%% file: minvar.tex
\documentclass[12pt,letterpaper]{article}

\input{config}

\input{header}

\input{format}

\input{macros}

\input{colors}

\begin{document}

\ifxetex
  \let\lsum\sum
  \renewcommand{\sum}{\bm{\lsum}} 

  \let\lprod\prod
  \renewcommand{\prod}{\bm{\lprod}}
\else
\fi

\title{\textbf{\Large The Dispersion Bias}}

\author{ Lisa Goldberg\footnote{Departments of Economics and Statistics and
Consortium for Data Analytics in Risk, University of California, Berkeley, CA
94720 and Aperio Group, {\tt lrg@berkeley.edu}.}~~Alex
Papanicolaou\footnote{Consortium for Data Analytics in Risk, University of
California, Berkeley, CA  94720, {\tt apapanicolaou@berkeley.edu}.}~~Alex
Shkolnik\footnote{Consortium for Data Analytics in Risk, University of
California, Berkeley, CA  94720, {\tt ads2@berkeley.edu}.} }

\date{November 4, 2017 \\ This Version: \today\footnote{
We thank the Center for Risk Management Research, the Consortium
for Data Analytics in Risk, and the Coleman Fung Chair for
financial support. We thank Marco Avellaneda, Bob Anderson, Kay
Giesecke, Nick Gunther, Guy Miller, George Papanicolaou, Yu-Ting
Tai, participants at the the 3rd Annual CDAR Symposium in
Berkeley, participants at the Swissquote Conference 2017 on
FinTech, and participants at the UC Santa Barbara Seminar in
Statistics and Applied Probability for discussion and comments.
We are grateful to Stephen Bianchi, whose incisive experiment
showing that it is errors in eigenvectors, and not in
eigenvalues, that corrupt  large minimum variance portfolios,
pointed us in a good direction. }}
\maketitle

\begin{abstract} 

\noindent Estimation error has plagued quantitative finance
since  Harry Markowitz launched modern portfolio theory in 1952.
Using random matrix theory, we characterize a source of bias in
the sample eigenvectors of financial covariance matrices.
Unchecked, the bias distorts weights of minimum variance
portfolios and leads to risk forecasts that are severely biased
downward. To address these issues, we develop an eigenvector
bias correction. Our approach is distinct from the
regularization and eigenvalue shrinkage methods found in the
literature. We provide theoretical guarantees on the improvement
our correction provides as well as estimation methods for
computing the optimal correction from data.

\end{abstract}

\newpage

\section{Introduction}
\label{S:intro}
\input{sections/intro}

\subsection{Our contributions}
\label{S:contributions}
\input{sections/contributions}

\subsection{Related literature}
\label{S:literature}
\input{sections/literature}


\section{Problem formulation}
\label{S:problem}
\input{sections/problem}

\subsection{Model specification \& assumptions}
\label{S:model}

\input{sections/model}

\subsection{PCA bias characterization}
\label{S:pca}
\input{sections/pca}

\subsection{Errors in optimized portfolios}
\label{S:metrics}
\input{sections/metrics}

\section{Bias correction}
\label{S:main}
\input{sections/main}

\section{Algorithm and extensions}
\label{S:algorithm}
\input{sections/algorithm}

\section{Numerical study}
\label{S:numerics}		
\input{sections/numerics}

\section{Summary}
\label{S:summary}
\input{sections/summary}
\newpage

\appendix

\section{Algorithm}
\label{A:algorithm}
\input{appendix/algorithm}

\newpage
\section{Tables}

\label{A:tables}
\input{tables/tables.tex}

\newpage

\section{Proof of main results}
\label{A:proof}
\input{appendix/proofs}

\section{Asymptotic estimates}
\label{A:lemma}
\input{appendix/asymptotics}

\bibliographystyle{agsm}
\bibliography{biblio}

\end{document}

%% file: config.tex
\usepackage[letterpaper]{geometry}
\usepackage{ifxetex}
\ifxetex
  \usepackage[T1]{fontenc}
  \usepackage[no-math]{fontspec}

  \usepackage[cal=euler, calscaled=1,
	      scr=boondox, scrscaled=1]{mathalfa}

  \let\hbar\relax
  \usepackage[mtpcal,mtphrb,mtpfrak,
  	      subscriptcorrection,zswash]{mtpro2}

  \DeclareMathSizes{12}{13}{9}{7}
\else
  \usepackage{lmodern}
  \usepackage{amsfonts}
  \usepackage{amssymb}
  \usepackage{mathrsfs}
  \newcommand{\mbf}{\mathbf} 
\fi

\usepackage{amsthm}

\newtheoremstyle{mytheoremstyle}
{\topsep}                    
{\topsep}                    
{\itshape}                   
{}                           
{\bfseries}                  
{.}                          
{.5em}                       
{}  

\theoremstyle{mytheoremstyle}

\makeatletter
\@ifclassloaded{report}{
   \newtheorem{theorem}{Theorem} }
{  \newtheorem{theorem}{Theorem}[section] }
\makeatother

\newtheorem{thm}[theorem]{Theorem}
\newtheorem{proposition}[theorem]{Proposition}
\newtheorem{lemma}[theorem]{Lemma}
\newtheorem{corollary}[theorem]{Corollary}
\newtheorem{definition}[theorem]{Definition}
\newtheorem{remark}[theorem]{Remark}

\newtheorem{condition}[theorem]{Condition}
\newtheorem{assumption}[theorem]{Assumption}

\newenvironment{prove}[1][\textsc{Proof.}]
 {\begin{proof}[#1] }
 {\end{proof} }

\usepackage{harvard}

%% file: header.tex
\usepackage{makecell}
\usepackage{graphicx}
\usepackage{enumerate}
\usepackage{framed}
\usepackage{multirow}
\usepackage{scrextend}
\usepackage{stackengine}
\usepackage{subcaption}
\usepackage{comment}
\usepackage{url}
\usepackage{bm}

\usepackage{marginnote}
\usepackage{setspace}

\makeatletter
\newcommand{\raisemath}[1]{\mathpalette{\raisem@th{#1}}}
\newcommand{\raisem@th}[3]{\raisebox{#1}{$#2#3$}}
\makeatother

\newif\ifhide
\hidetrue 

\usepackage{empheq}

\usepackage{pgffor}

\foreach \x in {A,...,Z}{%
  \expandafter\xdef\csname cal\x\endcsname{\noexpand %
	\ensuremath{\noexpand\mathcal{\x}}}
  \expandafter\xdef\csname scr\x\endcsname{\noexpand %
	\ensuremath{\noexpand\mathscr{\x}}}
  \expandafter\xdef\csname bb\x\endcsname{\noexpand %
	\ensuremath{\noexpand\mathbb{\x}}}
  \expandafter\xdef\csname rm\x\endcsname{\noexpand %
	\ensuremath{\noexpand\mathrm{\x}}}
  \expandafter\xdef\csname bf\x\endcsname{\noexpand %
	\ensuremath{\noexpand\mathbf{\x}}}
}

\newcommand{\wh}[1]{\widehat{#1}}

\newcommand{\hpt}{ \hspace{0.5pt} }

\newcommand{\pt}{ \hspace{1pt}  }
\newcommand{\ptt}{ \hspace{2pt} }

\newcommand{\cst}{ \hspace{0.5pt} : \hspace{0.5pt} }

\newcommand{\upto}{ \uparrow }

\newcommand{\tq}[1]{{\textquotedblleft #1\textquotedblright}}

\newcommand{\pa}[1]{\hspace{1pt} \left( \hspace{0.5pt} %
	{#1} \hspace{0.5pt} \right) \hspace{1pt}}

\newcommand{\p}[1]{ \hspace{1pt} ( \hspace{0.25pt} %
	{#1} \hspace{0.25pt} ) \hspace{1pt}}

\newcommand{\f}[1]{ \hspace{1pt} \{ \hspace{0.25pt} %
	{#1} \hspace{0.25pt} \} \hspace{1pt}}

\newcommand{\ff}[1]{ \hspace{1pt} \big\{ \hspace{0.5pt} %
	{#1} \hspace{0.5pt} \big\} \hspace{1pt}}

\makeatletter
\@ifpackageloaded{harvard}{}
  {\newcommand{\citeasnoun}{\cite}}
\makeatother

\newcommand{\ci}{\citeasnoun}
\newcommand{\cic}[2]{\citeasnoun[#1]{#2}}

\newcommand{\req}[1]{(\ref{#1})}

%% file: format.tex
\usepackage{harvard}
\usepackage{hyperref}
\hypersetup{colorlinks=true, allcolors=bamboo}

\usepackage{booktabs}

\usepackage{float}
\floatstyle{ruled}
\newfloat{algorithm}{hpt}{lop}
\floatname{algorithm}{Algorithm}

\usepackage{algorithm}
\usepackage{algpseudocode}

%% file: macros.tex
\newcommand{\cas}{\overset{a.s.}{\rightarrow}}

\newcommand{\sas}{\overset{a.s.}{\sim}}

\newcommand{\nv}{N}
\newcommand{\no}{T}

\newcommand{\pca}{PCA}

\newcommand{\mat}[1]{\mbf{#1}}
\newcommand{\bsig}{\bm{\Sigma}}
\newcommand{\blam}{\bm{\Lambda}}
\newcommand{\bdel}{\bm{\Delta}}
\newcommand{\bell}{\bm{L}}

\newcommand{\fv}{\ensuremath{\sigma^2}} 
\newcommand{\svol}{\ensuremath{\delta}}
\newcommand{\sv}{\ensuremath{\delta^2}} 
\newcommand{\fe}{\ensuremath{\beta}}    
\newcommand{\sfvol}{\mu}                
\newcommand{\sfv}{\sfvol^2}             
\newcommand{\sfve}{\hat{\sfvol}^2}
\newcommand{\sfvole}{\hat{\sfvol}}

\newcommand{\fee}{\ensuremath{\hat{\beta}}}
\newcommand{\fve}{\ensuremath{\hat{\sigma}^2}}
\newcommand{\fvole}{\ensuremath{\hat{\sigma}}}
\newcommand{\sve}{\ensuremath{\hat{\delta}^2}}
\newcommand{\svole}{\ensuremath{\hat{\delta}}}

\newcommand{\gam}[2]{\gamma_{#1,#2}}
\newcommand{\ang}[2]{\theta_{#1,#2}}

\newcommand{\err}{\scrE}

\newcommand{\minw}{\wh{w}}
\newcommand{\optw}{w_*}
\newcommand{\eqww}{w_e}

\newcommand{\rmtchisq}{\xi}
\newcommand{\rmtdeflate}{\Psi}

\newcommand{\eb}{\ensuremath{\hat{\beta}}}
\newcommand{\vs}{\ensuremath{\delta^2}}
\newcommand{\vf}{\ensuremath{\sigma^2}}

\newcommand{\vfe}{\ensuremath{\hat{\sigma}^2}}

\newcommand{\vse}{\ensuremath{\hat{\delta}^2}}

\newcommand{\bthr}{\beta_{\text{MV}}}
\newcommand{\ebthr}{\wh{\beta}_{\text{MV}}}

\newcommand{\te}{\scrT}

\newcommand{\mv}{{\rm MV}}
\newcommand{\vmf}{\ensuremath{\sigma^2_{m}}}

\newcommand{\ve}{\ensuremath{\sigma^2_\epsilon}}
\newcommand{\eve}{\ensuremath{\hat{\sigma}^2_\epsilon}}

\newcommand{\ven}{\ensuremath{\sigma^2_{\epsilon_n}}}

\newcommand{\hmv}{\widehat{\mv}}

\newcommand{\vmve}{\ensuremath{\sigma^2_{\widehat\mv}}}
\newcommand{\evmve}{\ensuremath{\hat{\sigma}^2_{\widehat\mv}}}

\newcommand{\wmve}{\ensuremath{w_{\widehat\mv}}}
\newcommand{\pvfr}{\text{$\scrR_{\widehat\mv}$}}

\newcommand{\ep}{\epsilon}

\newcommand{\lrgtext}[1]{\oystertext{#1}}
\newcommand{\adstext}[1]{\bambootext{#1}}

\newcommand{\lrglmn}[1]{{\reversemarginpar \setstretch{0.64} %
	\marginnote{{\scriptsize 
	\oystertext{\emph{#1}--lisa}}}} }
\newcommand{\lrgrmn}[1]{{\normalmarginpar \setstretch{0.64} %
	\marginnote{{\scriptsize 
	\oystertext{\emph{#1}--lisa}}}} }

\newcommand{\adslmn}[1]{{\reversemarginpar \setstretch{0.64} %
	\marginnote{{\scriptsize 
	\bambootext{\emph{#1}--alex.s}}}} }

%% file: colors.tex
\usepackage{color}
\usepackage[table]{xcolor}

\definecolor{red}{rgb}{1,0,0}
\definecolor{gray}{rgb}{0.5,0.5,0.5}
\definecolor{darkgray}{rgb}{0.4,0.4,0.4}
\definecolor{blue}{rgb}{0,0,1}
\definecolor{green}{rgb}{0,1,0}

\definecolor{deluge}{RGB}{124, 113, 173}
\definecolor{bamboo}{RGB}{220, 92, 5}
\definecolor{yellow}{RGB}{255, 172, 0}
\definecolor{orange}{RGB}{255, 144, 0}
\definecolor{oyster}{RGB}{151, 139, 125}
\definecolor{coral}{RGB}{199, 186, 167}
\definecolor{downy}{RGB}{110, 197, 184}

\def\bambootext#1{{\color{bamboo}{{#1}}\color{bamboo}}}

\def\oystertext#1{{\color{oyster}{{#1}}\color{oyster}}}

%% file: sections/intro.tex

Harry Markowitz transformed finance  in 1952 by framing
portfolio construction as a tradeoff between mean and variance
of return.  This application of mean-variance optimization is
the basis of theoretical  breakthroughs as fundamental as the
Capital Asset Pricing Model (CAPM) and Arbitrage Pricing Theory
(APT), as well as practical innovations as impactful as Exchange
Traded Funds.\footnote{ The seminal paper is \ci{markowitz1952}.
See \ci{treynor1962} and \ci{sharpe1964} for the Capital Asset
Pricing Model and \ci{ross1976} for the Arbitrage Pricing
Theory.} Still, all financial applications of mean-variance
optimization suffer from estimation error in covariance
matrices, and we highlight two difficulties.  First, a portfolio
that is optimized using an estimated covariance matrix is never
the true Markowitz portfolio. Second, in current practice, the
forecasted risk of the optimized portfolio is typically too low,
sometimes by a wide margin. Thus, investors end up with the
wrong portfolio, one that is riskier, perhaps a lot riskier,
than anticipated.

In this article, we address these difficulties by correcting a
systematic bias in the first eigenvector of a sample covariance
matrix. Our setting is that of a typical factor
model,\footnote{More precisely, the eigenvalues of the
covariance matrix corresponding to the factors grow linearly in
the dimension. This is not the traditional random matrix theory
setting in which all eigenvalues are bounded, nor that of
\tq{weak} factors, e.g., \ci{onatski2012}.} but our statistical
setup differs from most recent literature. In the last two
decades, theoretical and empirical emphasis has been on the case
when the number of assets $\nv$ and number of observations $\no$
are both large. In this regime, consistency of principal
component analysis (\pca) estimates may be established
\cite{bai2008}.  Motivated by many  applications, we
consider the setting of relatively few observations (in
asymptotic theory: $\nv$ grows and $\no$ is fixed).  Indeed, an
investor often has a portfolio of thousands of securities
but only hundreds of observations.\footnote{While high frequency
data are available in some markets, many securities are observed
only at a daily horizon or less frequently. Moreover, markets
are non-stationary, so even when there is a long history of data
available, its relevance to some problems is questionable.}
{\pca} is applied in this environment in the early, pioneering
work by \ci{connor1986} and \ci{connor1988}, but also very
recently \cite{wang2017}. In this high dimension, low
sample-size regime, {\pca} factor estimates necessarily carry a
finite-sample bias. This bias is further amplified by the 
optimization procedure that is required to compute a Markowitz 
portfolio.

An elementary simulation experiment reveals that in a large
minimum variance portfolio, errors in portfolio weights are
driven by the first principal component, not its
variance.\footnote{This experiment was first communicated to us
by Stephen Bianchi.} The fact that the eigenvalues of the sample
covariance matrix are not important requires some nontrivial
analysis, which we carry out.  In particular, we show (in our
asymptotic regime) that the bias in the dominant sample
eigenvalue does not effect the performance of the estimated
minimum variance portfolio. Only the bias in the dominant sample
eigenvector needs to be addressed. We measure portfolio
performance using two well-established metrics. Tracking error,
the workhorse of financial practitioners, measures deviations in
weights between the estimated (optimized) and optimal
portfolios. We use the variance forecast ratio, familiar to both
academics and practitioners,  to measure the accuracy of the
risk forecast of the portfolio, however right or wrong that
portfolio may be.

To develop some intuition for the results to come, consider a
simplistic world where all security exposures to the dominant
(market) factor are identical.  With probability one, a {\pca}
estimate of our idealized, dominant factor will have higher
dispersion (variation in its entries).  Decreasing this
dispersion, obviously, mitigates the estimation error. We denote
our idealized, dominant factor by~$z$.  We prove that the same
argument applies to any other dominant factor along the
direction of $z$ with high probablity for $\nv$ large.  Thus
moving our {\pca} estimate towards $z$, by some amount, is very
likely to decrease estimation error. In the limit ($\nv \upto
\infty$), the estimation error is reduced with probability one.
The larger the component of the true dominant factor along $z$
is, the more we can decrease the estimation error.

While a careful proof of our result relies on some recent theory
on sample eigenvector asymptotics, rule of thumb versions have
been known to practitioners since the 1970s (see footnote
\ref{blumebeta} and the corresponding discussion).  Indeed, the
dominant risk factor consistently found in the US and many other
developed public equity markets has most (if not all) individual
equities positively exposed to it. In other words, empirically,
the dominant risk factor has a significant component in $z$.
Our characterization of the dispersion bias may then be viewed
as a formalization of standard operation procedure.

The remainder of the introduction discusses our contributions
and the related literature.  Section \ref{S:problem} describes
the problem and fundamental results around the sample covariance
matrix and PCA.  In Section \ref{S:main}, we present our main
results on producing a bias corrected covariance estimate.
Section \ref{S:algorithm} discusses the implementation of our
correction for obtaining data-driven estimates.  Finally, in
Section \ref{S:numerics} we present numerical results
illustrating the performance of our method in improving the
estimated portfolio and risk forecasts.

%% file: sections/contributions.tex
We contribute to the literature by providing a method that
significantly improves the performance of {\pca}-estimated
minimum-variance portfolios. Our approach and perspective appear
to be new. We summarize some of the main points.

Several authors (see above) have noted that sample eigenvectors
carry a bias in the statistical and model setting we adopt. We
contribute in this direction by, first, recognizing that it is
the bias in the first sample eigenvector that drives the
performance of {\pca}-based, minimum-variance portfolios.
Second, we show that this bias may in fact be corrected to some
degree (cf. discussion below (3.7) in \ci{wang2017}). In our
domain of application this degree is material.  We point out
that eigenvalue bias, which has been the focus in most
literature, does not have a material impact on minimum-variance
portfolio performance. This motivates lines of research into
more general Markowitz optimization problems. Finally, our
correction can be framed geometrically in terms of the spherical
law of cosines.  This perspective illuminates possible
extensions of our work. We discuss this further in our
concluding remarks.

We also develop a bias correction and show that it outperforms
standard {\pca}. Minimum variance portfolios constructed with
our corrected covariance matrix are materially closer to
optimal, and their risk forecasts are materially more accurate.
In an idealized one-factor setting, we provide theoretical
guarantees for the size of the improvement.  Our theory also
identifies some limitations. We demonstrate the efficacy of the
method with an entirely data-driven correction.  In an
empirically calibrated simulation, its performance is far closer
to the theoretically optimal than to standard {\pca}.

%% file: sections/literature.tex

The impact of estimation error on optimized portfolios has been
investigated thoroughly in simulation and emprical settings. For
example, see \ci{jobson1980}, \ci{britten1999}, \ci{bianchi2017}
and the references therein. \ci{demiguel2007} compare a variety
of methods for mitigating estimation error, benchmarking againt
the equally weigted portfolio in out-of-sample tests. They
conclude that unreasonably long estimation windows are required
for current methods to consistently outperform the benchmark.
We review some methods most closely related to our
approach.\footnote{The literature on this topic is extensive.
We briefly mention a few important references that do not
overlap at all with out work. \ci{michaud2008} recommends the
use of bootstap resampling. \ci{lai2011} reformulate the
Markowitz problem as one of stochastic optimization with unknown
moments. \ci{goldfarb2003} develop a robust optimization
procedure for the Markowitz problem by embedding a factor
structure in the constraint set.}

Early work on estimation error and the Markowitz problem was
focused on Bayesian approaches. \ci{vasicek1973} and
\ci{frost1986} were perhaps the first to impose informative
priors on the model parameters.\footnote{Preceeding work
analyzed diffuse priors and was shown to be inneficient
\cite{frost1986}. The latter, instead, presumes all stocks are
identical and have the same correlations. \ci{vasicek1973}
specified a normal prior on the cross-sectional market betas
(dominant factor).} More realistic priors incorporating
multi-factor modeling are analyzed in \ci{pastor2000} 
(sample mean) and \ci{gillen2014} (sample covariance).
Formulae for Bayes' estimates of the return mean and
covariance matrix based on normal and inverted Wishart priors
may be found in \cic{Chapter 4, Section 4.4.1}{lai2008}.

A related approach to the Bayesian framework is that of
shrinkage or regularization of the sample covariance
matrix.\footnote{In the Bayesian setup, sample estimates are
\tq{shrunk} toward the prior \cite{lai2008}.} Shrinkage methods
have been proposed in contexts where little underlying structure
is present \cite{bickel2008a} as well as those in which a factor
or other correlation structure is presumed to exist (e.g.
\ci{ledoit2003}, \ci{ledoit2004}, \ci{fan2013} and
\ci{bun2016}). Perhaps surprisingly, shrinkage methods turn out
to be related to placing constraints on the portfolio weights in
the Markowitz optimization. \ci{jagannathan2003} show
that imposing a positivity constraint typically shrinks the
large entries of the sample covariance downward.\footnote{This
is generalized and analyzed further in
\ci{demiguel2009}.} 

As already mentioned, factor analysis and PCA in particular
play a prominent role in the literature.
It appears that while eigenvector
bias is acknowledged, direct\footnote{Several approaches to 
alter the sample eigenvectors indirectly (e.g. shrinking 
the sample towards some structured covariance) do exist. 
However, the analysis of these approaches is not focused on 
characterizing the bias inherent to the sample eigenvectors
themselves.}  
 bias corrections 
are made only to the eigenvalues corresponding to the principal
components (e.g. \ci{ledoit2011} and \ci{wang2017}).
Some work on characterizing the behavior of
sample eigenvectors may be found in \ci{paul2007} and 
\ci{shen2016}.
In the setting of Markowitz portfolios, the
impact of eigenvalue bias and optimal corrections are
investigated in in \ci{el2010} and \ci{el2013}.

Our approach also builds upon several profound contributions in
the literature on portfolio composition.  In an influential
paper, \ci{green1992} observe the importance of the structure of
the dominant factor to the composition of minimum variance
portfolios. In particular, the \tq{dispersion} of the dominant
factor exposures drives the extreme positions in the portfolio
composition. This dispersion is further amplified by estimation
error, as pointed out in earlier work by \ci{blume1975} (see
also \ci{vasicek1973}).  These early efforts have led to a
number of heuristics\footnote{For example, the Blume and Vasicek
(beta) adjustments. See the discussion of Exhibit 3 and footnote
7 in \ci{clarke2011}.\label{blumebeta}} to correct the sample
bias of dominant factor estimates.

%% file: sections/problem.tex
We address the impact of estimation error on Markowitz
portfolio optimization. To streamline the exposition, we restrict
our focus to the minumum variance 
portfilio. In particular, given a covariance matrix 
$\wh{\bsig}$ estimated from $\no$ observations of returns to 
$\nv$ securities, we consider the optimization problem,
\begin{equation} \label{minvar}
\begin{aligned} 
  &\min_{w \in \bbR^{\nv}} w^\top \wh{\bsig} w
  \\ &\hspace{8pt} w^\top 1_{\nv} = 1 \pt .
\end{aligned}
\end{equation} 
We denote by $\minw$ the solution to $\req{minvar}$, the
estimated minimum variance portfolio. Throughout, $1_\nv$ is
the $\nv$-vector of all ones. It is well-known that the portfolio
weights, $\minw$, are extremely sensitive to errors in the
estimated model, and risk forecasts for the optimized portfolio
tend to be too low.\footnote{Extreme sensitivity of portfolio
weights to estimation error and the downward bias of risk
forecasts are also found in the optimized portfolios constructed
by asset managers.  Portfolio specific corrections of the
dispersion bias discussed in this article are  useful in
addressing these practical problems.  The focus on the global
minimum variance portfolio in this article highlights the
essential logic of our analysis in the simplest possible
setting.} We aim to address these issues in a high-dimension,
low sample-size regime, i.e., $T\ll N$.
 
We are also interested in the equally weigthed portfolio where
$\eqww = 1_{\nv}/\nv$, a very simple non-optimized portfolio
frequenly employed as a benchmark. We use this benchmark to test
whether the corrections we make for improving the minimum
variance portfolio are not offset by degraded performance
elsewhere.


%% file: sections/model.tex
We consider a one-factor, linear model for returns to $\nv \in
\bbN$ securities.  Here, a generating process for the $N$-vector
of excess returns $R$ takes the form
\begin{align} 
  R =  \phi \beta + \ep \label{1factor}
\end{align}
where $\phi$ is the return to the factor, $\beta = \p{\beta_1,
\dots, \beta_{\nv}}$ is the vector of factor exposures, and $\ep
= \p{\ep^1, \dots, \ep^{\nv}}$ is the vector of diversifiable
specific  returns.  While the returns $\p{\phi,\ep} \in \bbR
\times \bbR^{\nv}$ are random, we treat each exposure $\beta_n
\in \bbR$ as a constant to be estimated.  Assuming $\phi$ and
the $\f{\ep_n}$ are mean zero and pairwise uncorrelated, the
$\nv\times \nv$ covariance matrix of $R$ can be expressed as
\begin{align} \label{cov}
  \bsig = \vf \beta \beta^\top + \bdel \pt .
\end{align}
Here, $\sigma^2$ is the variance of $\phi$ and $\bdel$ a
diagonal matrix with $n$th entry $\bdel_{nn} = \delta^2_n$, the
variance of $\ep_n$. Estimation of $\bsig$ is central to
numerous applications.

We consider a setting in which $\no$ observations $\f{R_t}_{t =
1}^{\no}$ of the vector $R$ are generated by a latent
time-series $\f{\phi_t, \ep_t}_{t=1}^{\no}$ of $\p{\phi,\ep}$.
It is standard to assume  the observations are i.i.d. and we do
so throughout.  Finite-sample error distorts measurement of the
parameters $\p{\vf\hspace{-4pt}, \beta, \delta^2}$ leading
to the estimate, 
\begin{align} \label{ecov} 
\wh{\bsig} = \vfe \eb \eb^\top + \wh{\bdel} \pt , 
\end{align} 
which approximates $\req{cov}$ by using the estimated model
$\p{\vfe\hspace{-4pt}, \eb, \vse}$.

Without loss of generality we assume the following condition 
throughout.
\begin{condition}
Both $\beta$ and any estimate $\eb$ are normalized as 
$\|\beta\| = \|\eb\| = 1$.
\end{condition}

We require further statistical and regularity assumptions
on our model for our techical results.  
These conditions stem from our use of recent
work on spiked covariance models \cite{wang2017}. Some may 
be relaxed in various ways. 
Our numerical results (Section \ref{S:numerics}) 
investigate a much more general setup.

\begin{assumption}\label{a:vol}
  The factor variance $\vf = \sigma_{\nv}^2$  satisfies
  $\frac{\nv}{\no\sigma_{\nv}^2} \rightarrow c_1$ as 
  $\nv \upto \infty$ for fixed integer $\no \ge 1$ and  
  $c_1 \in (0,\infty)$. Also, $\bdel = \delta^2 \mat{I}$ 
  for fixed $\delta \in (0,\infty)$.
\end{assumption}

\begin{assumption}\label{a:gauss}
  The returns $\f{R_t}_{t=1}^{\no}$ are i.i.d.
  with $R_1 \sim \mathcal{N}(0, \bsig)$.
\end{assumption}


\begin{assumption}\label{a:orient}
  For $z = 1_N/\sqrt{N}$ we have $\sup_\nv \gam{\fe}{z}
 < 1$ and $\sup_\nv \gam{\fee}{z} < 1$.
  Also, $\fe$ and $\fee$ are 
  oriented in such a way that $\beta^\top z$ and 
$\eb\top z$ are nonnegative.
\end{assumption}

The requirement in Assumption \ref{a:vol} that
the factor variance grow in dimension while the specific
variance stays bounded is pervasive in the factor modeling
literature \cite{bai2008}.  The extra requirement that the
specific risk is homogenous (i.e. $\bdel$ is a scalar matrix) is
restrictive but shortens the technical proofs significanly.  It
is also commonplace to the spiked covariance model
literature.\footnote{In particular, our need for this condition
stem from our use of results from \ci{shen2016}. The assumption
may be relaxed by imposing regularity conditions on the entries
of $\bdel$ at the expense of a more cumbersome exposition. It
may also be removed entirely if we consider the regime $\nv,\no
\upto \infty$ as is more common in the literature. We do not
pursue this because many important pratical applications are
restricted to only a small number of observations.} We discuss
the adjustments required for heterogenous specific risk (i.e.
$\bdel$ diagonal) in Section \ref{S:heterogenous}.  The
distributional Assumption~\ref{a:gauss}
facilitates several steps in the proofs. In particular, it
allows for an elegant characterization of the systematic bias in
PCA, i.e., the bias in the first eigenvector of the sample
covariance matrix of the returns $\f{R_t}$ (see Section
\ref{S:pca}).  Assumption~\ref{a:orient} is not
much of a restriction. First, all results can be easily extended
to the case $\beta = z$, which is simply a point of
singularity and thus requires a separate treatment. The
orientation requirement is essentially without loss of 
generality. We will see (in Section \ref{S:main}) that the 
vector $z$ plays a special role in our bias correction
procedure. And if $\beta\top z < 0$, we would simply consider
$-z$.

%% file: sections/pca.tex

We consider {\pca} as the starting point for our analysis as its
use  for risk factor identification is widespread in the
literature. It is appropriate when $\vf$ is much larger than
$\|\bdel\|$ (e.g., Assumption \ref{a:vol}). Assembling a data
matrix $\mat{R} = \p{R_1, \dots, R_T}$, we will denote the
(data) sample covariance matrix by $\mat{S} = \no^{-1} \mat{R}
\mat{R}^\top$. {\pca} identifies $\vfe$ with the largest
eigenvalue of $\mat{S}$ and $\eb$ with the corresponding
eigenvector (the first principal component).  The diagonal
matrix of specific risks $\bdel$ is estimated as $\wh{\bdel} =
\text{\bf diag} \p{ \mat{S} - \vfe \eb \eb^\top}$, which
corresponds to least-squares regression of $\mat{R}$ onto the
estimated factors. Finally, the estimate $\wh{\bsig}$ of the
covariance $\bsig$ is assembled as in~$\req{ecov}$.

Bias in the basic {\pca} estimator above arises from the use of
the  sample\footnote{If $\bsig$ replaces $\mat{S}$,
sample bias vanishes and the estimator is asymptotically exact
as $N \upto \infty$.} covariance matrix $\mat{S}$. 
We focus on the high-dimension, low sample size regime which is
most appropriate for the practical applications we consider.
Asymptotically, $\no$ is fixed and $\nv \upto \infty$, a regime
in which the {\pca} estimates of $\vf$ and $\beta$ are 
not consistent. We summarize some recent results from the 
literature below. We also
state our characterization of PCA bias as it pertains 
to its pricipal components. Our result is novel in that
it suggests a remedy.


\begin{figure}[htp]
\centering
\begin{subfigure}[t]{.4\textwidth}
\centering
\includegraphics[width=\linewidth]{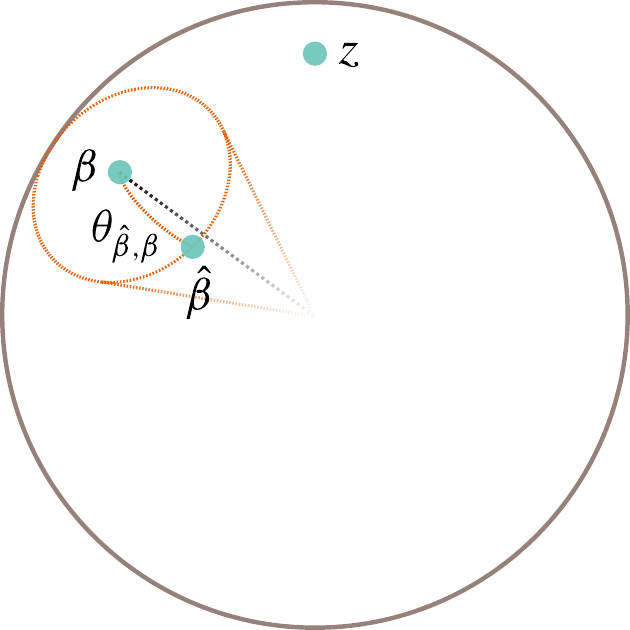}
 \caption{$\eb$ lies near the cone defined by the
$\rmtdeflate$ in $\req{ang-error}$ w.h.p. for large $\nv$. 
There is no reference frame to detect the bias since $\beta$ 
is unknown.}\label{fig:geom}
\end{subfigure}
\hspace{0.32in}
\begin{subfigure}[t]{.4\textwidth}
\centering
\includegraphics[width=\linewidth]{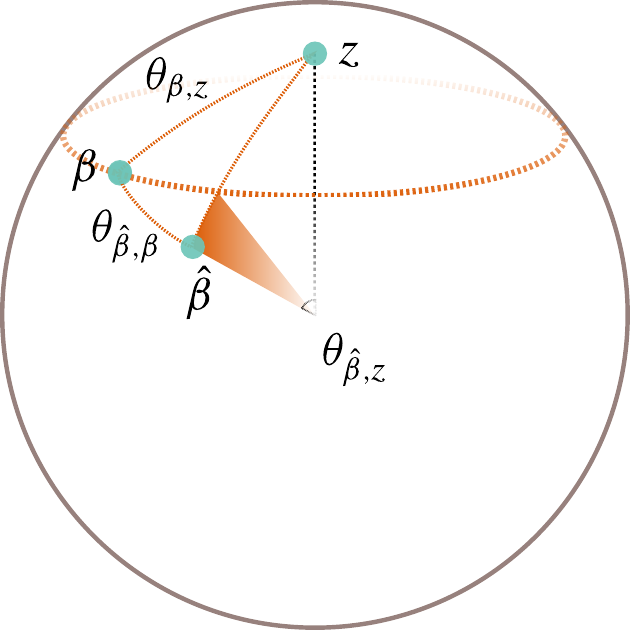}
 \caption{$\theta_{\hat\beta,z} > \theta_{\beta,z}$ w.h.p.
for large~$\nv$ (their difference is shaded). 
{\pca} estimates exhibit a systematic (dispersion) bias 
relative to vector $z$.}\label{fig:bias}
\end{subfigure}
\caption{A unit sphere with example vectors $\beta$,
$\eb$ and $z = 1_\nv/\sqrt{\nv}$. The vector $z$ 
provides a reference frame in which {\pca} bias may be 
identified. Note, the angle $\theta_{x,y}$ is also the 
arc-length between points $x$ and $y$ on the unit sphere.}
\label{F:bias}
\end{figure}

Let $\ang{\fee}{\fe}$ denote the angle between $\beta$ and its
{\pca} estimate $\fee$. 
\ci{shen2016} showed, under Assumption~\ref{a:vol} and mild
conditions on the moments of $\f{R_t}$, that there exist random
variables $\rmtdeflate > 1$ and $\rmtchisq \ge 0$ such 
that\footnote{More precisely, $\rmtdeflate^2 = 1 + \vs c_1
/\rmtchisq$ where $c_1$ is identified in Assumption
\ref{a:vol}.}
\begin{align}
   \cos \ang{\fee}{\fe} \ptt &\cas  \ptt \rmtdeflate^{-1}
   \label{ang-error}  \\
   \fve / \fv \ptt 
   &\cas \ptt \rmtdeflate^{2} \rmtchisq 
  \label{vf-error} 
\end{align}
as $\nv \upto \infty$  where $\rmtchisq$ and $\rmtdeflate$
depend on $\no$ (fixed).  The pair $\p{\rmtchisq, \rmtdeflate}$
characterizes the error in the {\pca} estimated model
asymptotically.  As the sample size $\no$ increases, both
$\rmtchisq$ and $\rmtdeflate$ approach one almost surely, i.e.,
{\pca} supplies consistent estimates. Since $\rmtdeflate > 1$,
the estimate $\fve$ tends to be biased upward (whenever
$\rmtchisq$ fluctuates around one).  Under Assumption
\ref{a:gauss}, $\rmtchisq = \chi^2_\no / \no$ where $\chi^2_\no$
has the chi-square distribution  with $\no$ degrees of freedom.
Here, $\rmtchisq$ is concentrated around one with high
probability (w.h.p.) for even moderate values of $\no$.

Identifying a (systematic) bias in {\pca} estimates of $\beta$
requires a more subtle analysis. Observe from $\req{ang-error}$
that  $\rmtdeflate$ defines a cone near which the estimate $\eb$
will lie with high probability for large $N$ (see panel (a) of
Figure \ref{F:bias}).  However, this does not point to a
systematic error that can be corrected.\footnote{Recently,
\ci{wang2017} provided CLT-type results for the asymptotic
distribution of sample eigenvectors (Theorem 3.2).  They remark
on the near \tq{impossibility} of correcting sample eigenvector
bias. This follows from their choice of coordinate system (cf.
Figure \ref{F:bias}).} Indeed, it is not apriori clear where on
the cone the vector $\eb$ resides as $\req{ang-error}$ provides
information only about the angle $\ang{\fee}{\fe}$ away from the
unknown vector $\beta$. We provide a result (see Theorem
\ref{T:bias} below) that sheds light on this problem.

Recall the vector $z = 1_\nv/\sqrt{\nv}$. We consider, 
not the angle $\ang{\fee}{\fe}$ between $\beta$ and $\eb$, but
the their respective angles $\theta_{\beta,z}$ and
$\theta_{\eb,z}$ to this reference vector. 

\begin{theorem}[PCA bias] \label{T:bias}
Suppose that Assumptions \ref{a:vol}, \ref{a:gauss} and
\ref{a:orient} hold and let $\eb$ be a {\pca} estimate of
$\fe$. Then, $\cos \ang{\fe}{z} \sas 
\rmtdeflate \cos\ang{\fee}{z}$ as $\nv \upto \infty$ and, 
in particular, we have that $\theta_{\eb,z}$
exceeds $\theta_{\beta,z}$ with high probability for large
$\nv$. 
\end{theorem}

The proof of this result is deferred to Appendix \ref{A:proof}. 
It applies the spiked covariance model results of 
\ci{shen2016} and \ci{paul2007} on sample eigenvectors 
using a decomposition in terms of the reference vector $z$.

%% file: sections/metrics.tex

Estimation error causes two types of difficulties in optimized
portfolios. It distorts portfolio weights, and  it biases the
risk of optimized portfolios downward. Both effects are present
for the minimum variance portfolio $\minw$, constructed as the
solution to $\req{minvar}$ using some estimate $\wh{\bsig}$ of
the returns covariance $\bsig$. We now define the metrics for
assessing the magnitude of these two errors.


We denote by $\optw$ the optimal portfolio, i.e., the solution
of $\req{minvar}$ with $\wh{\bsig}$ replaced by $\bsig$.  Since
the latter is positive definite, the optimal portfolio weights
$\optw$ may be given explicitly.\footnote{Indeed, the
Sherman-Morrison-Woodbury formula yields an explicit solution
even for a multi-factor model and with a guaranteed mean return
\cite{karoui2008}. In our setting,
\begin{align}
  \hspace{16pt}
  \optw \propto 
  \bdel^{-1} \p{1_{\nv} \bthr - \fe} \hpt, 
  \hspace{16pt} 
  \bthr = \frac{1 + \vf \sum_{k=1}^{\nv} {\beta^2_k}/{\vs_k}}
  { \vf \sum_{k=1}^{\nv} \beta_k/\vs_k} \pt .
\label{eq:clarke}
\end{align}}
We define,
\begin{align}\label{te}
  \te_{\minw}^2  = \left(\optw - \minw\right)^\top 
  \bsig \left(\optw - \minw\right) \pt, 
\end{align}
the (squared) tracking error of $\wh w$. Here, $\te^2_{\minw}$
measures the distance between the optimal and estimated
portfolios, $\optw$ and $\minw$. Specifically, it is the square of
the width of the distribution of return differences 
$\optw -\minw$.

The variance of portfolio $\minw$ is given by
$\minw^\top\wh{\bsig}\minw$ and its true variance is
$\minw^\top{\bsig}\minw$. We define,
\begin{align} \label{var-ratio}
  \scrR_{\wh w} = \frac{ \minw^\top \wh{\bsig} \minw}
  { \minw^\top \bsig \minw} \pt, 
\end{align}
the variance forecast ratio. Ratio $\req{var-ratio}$ is less
than one when the risk of the portfolio $\minw$ is
underforecast.\footnote{With respect to the equally weighted 
portfolio, $\eqww$, (the tracking error of which is zero)
we only consider the variance forecast ratio.}



Metrics $\req{te}$ and $\req{var-ratio}$ quantify the errors in
portfolio weights and risk forecasts induced by estimation
error.\footnote{For a relationship to more standard error norms
see \ci{wang2017}.} 
We analyze $\te_{\minw}$ and $\scrR_{\minw}$ asymptotically.
Again recall $z = 1_\nv/\sqrt{\nv}$ and let $\gam{x}{y}
= x^\top y$. Note, $\gam{x}{y} = \cos \ang{x}{y}$ 
whenever $x$ and $y$ lie on the surface of a 
unit sphere as in Figure \ref{F:bias}. Define,
\begin{align} \label{error}
  \err = \frac{\gam{\fe}{z} - \gam{\fe}{\fee} \gam{\fee}{z}}
  { \sin^2 \ang{\fee}{z}} \pt .
\end{align}
The variable $\err$ drives the
asymptotics of our error metrics. Note that $\err = 0$
when $\fee = \fe$. Indeed, it is not difficult to show 
that for any estimates $\p{\fvole, \fee, \svole}$ satisfying
Assumption \ref{a:orient}, if
$\err$ is bounded away from zero (i.e. $\inf_\nv \err > 0$),
\begin{equation} \label{eq:gen-error}
\hspace{0.32in} 
\begin{aligned}
  \te^2_{\wh w} &\sim \sfv \err^2 \\
  \scrR_{\wh w} &\sim
    \frac{ \nv^{-1} \sve }
    {\sfv \err^2 \sin^2 \ang{\fee}{z} }
\end{aligned}
\hspace{0.32in} \p{\nv \upto \infty}.
\end{equation}
where $\sfv = \fv/\nv$ which is bounded in $\nv$.
This result (with our all assumptions above relaxed) is 
given by Proposition \ref{P:asymptotics} of Appendix
\ref{A:lemma}.

\begin{corollary}[PCA performance]\label{thm:pca_bias}
Suppose Assumptions \ref{a:vol}, \ref{a:gauss} and \ref{a:orient}
hold. For the {\pca}-estimator of the minimum variance portfolio
$\minw$, the tracking error squared 
$\te^2_{\minw}$ is bounded away from zero  
and variance forecast ratio $\scrR_{\minw}$ tends to zero
as $\nv \upto \infty$. In particular, for the {\pca}
estimator, the 
error $\err$ in \eqref{error} satisfies 
\begin{align} \label{pca-error}
  \err \sas \gam{\fe}{z} 
  \pa{ \frac{1 - \rmtdeflate^{-2}}
 {1 -\rmtdeflate^{-2} \gam{\fe}{z}^2}}
\end{align}
as $\nv \upto \infty$ where $\rmtdeflate>1$ 
is the random variable in \eqref{ang-error}. 
\end{corollary}
\begin{prove} Expression $\req{pca-error}$ follows from
\eqref{ang-error},
Theorem $\ref{T:bias}$ (i.e. $\gam{\fe}{z} \sas \rmtdeflate
\gam{\fee}{z}$) and the identity $1 - \gam{\fee}{z}^2 = \sin^2
\ang{\fee}{z}$. The remaining claims follow from 
\eqref{eq:gen-error}.
\end{prove}

The result states that the variance forecast ratio for the 
minimum variance portfolio $\minw$ that uses the {\pca} 
estimator of Section \ref{S:pca} is asymptotically $0$. The
estimated risk will be increasingly optimistic as $\nv
\rightarrow \infty$. This is entirely due to estimation
error between the sample eigenvector and the population
eigenvector.  As $\nv$ grows, the forecast risk becomes
negligible relative to the true risk, rendering the PCA
estimated minimum variance portfolio worse and worse. The
tracking error is also driven by the error of the sample
eigenvector and for increasing dimension, its proximity to the
true minimum variance portfolio as measured by tracking error is
asymptotically bounded below (away from zero).

%% file: sections/main.tex

Our Theorem \ref{T:bias} characterizes the bias of the  
{\pca} estimator in terms of the vector~$z$. This is the unique
(up to negation) dispersionless vector on the unit sphere,
i.e., its entries do not vary. Of course when $z = \beta$, then
the its {\pca} estimate $\eb$ will have higher dispersion with
probability one. The argument works along the projection of 
any $\beta$ along $z$, given by $\gamma_{\beta,z}$. Our 
{\pca} bias characterization implies that $\gamma_{\beta,z}
> \gamma_{\eb,z}$, or equivalently, 
$\theta_{\eb,z} > \theta_{\beta,z}$.
with high probability (for large $\nv$). Figure \ref{fig:bias}
illustrates this systematic {\pca} bias and clearly suggests
a correction: a shrinkage of the {\pca} estimate $\eb$
towards~$z$.

\subsection{Intuition for the correction}
\label{S:shrink}
\input{sections/shrink}

\subsection{Statement of the main theorem}
\label{S:theorem}

As noted above, we can improve tracking error and variance
forecast ratio by reducing the angle between the estimated
eigenvector  and the true underlying eigenvector, $\beta$,
equivalently, by replacing $\eb$ with an appropriate choice of
$\eb_\rho$.  We find an optimal value $\rho^*_{\nv}$ for a
particular $\nv$.  We present our method and its impact on
tracking error and variance forecast ratio for a minimum
variance portfolio below.  We restrict to the case of
homogeneous specific risk where $\bdel = \delta^2 \mat{I}$ for
expositional purposes but consider the full- fledged case in
empirical results in Section \ref{S:numerics}.

In the following theorem, we also provide a correction to the
sample eigenvalue.  Our bias correction for the sample
eigenvector introduces a bias in the variance forecast ratio for
the equally weighted portfolio.  We shrink the sample
eigenvalue, treated as the variance of the estimated factor, to
debias the variance forecast ratio for the equally weighted
portfolio.

\begin{thm}\label{thm:main}

Suppose Assumptions \ref{a:vol},
\ref{a:gauss}, \ref{a:orient}
hold and 
denote by $\hat{\delta}$ any
estimator of the specific risk $\delta$.
Define the oracle corrected estimate by $\eb_{\rho^*} :=
\eb_{\rho^*_{\nv}}$ where the finite sample (fixed $\nv$ and
$\no$) optimal value $\rho^*_{\nv}$ solves 
the equation $0 = r_{\eb_{\rho}} - \gamma_{\beta, \eb_{\rho}}$ 
(or equivalently, maximizes $\gamma_{\beta, \eb_{\rho}}$).

  \begin{enumerate}[i.]

    \item The finite sample optimal oracle $\rho^*_{\nv}$ 
     is given by,
      \begin{align}\label{eq:rhostar}
        \rho^*_{\nv}
         = \frac{\gamma_{\beta, z} - \gamma_{\beta,\eb} \gamma_{\eb, z}}
            {\gamma_{\beta,\eb} - \gamma_{\beta, z} \gamma_{\eb, z}}.
      \end{align}

    \item For the oracle value $\rho^*_{\nv}$, the tracking
      error and forecast variance ratio for the minimum variance portfolio for
      $\wh{\bsig}_{\rho^*_{\nv}} = \vfe \eb_{\rho^*}\eb_{\rho^*}^T +
      \hat{\delta}^2 \mat{I}$ satisfy,
      \begin{gather}
        \te^2_{\wh w} \sas \frac{1}{\nv}\delta^2
            \frac{ \gamma_{\eb_{\rho^*}, z}^2 - \gamma_{\beta, z}^2}
              {(1 - \gamma_{\eb_{\rho^*}, z}^2)(1 - \gamma_{\beta, z}^2)} \\
        \scrR_{\wh w} \sas { \hat{\delta}^2 }/{ {\delta}^2 }.
      \end{gather}
      That is, after the optimal correction, the forecast variance ratio for the
      minimum variance portfolio no longer converges to 0 while the tracking
      error to the true minimum variance portfolio does.

    \item For $\no$ fixed, $\nv \rightarrow \infty$, we have,
     $\rho^*_{\nv} \sas \bar{\rho}_\nv$ where 
      \begin{align}\label{eq:rhostar_asy}
          \bar{\rho}_\nv 
            = \frac{ \gamma_{\beta, z} }{ 1 - \gamma_{\beta, z}^2}
                \left(\rmtdeflate - \rmtdeflate^{-1}\right).
      \end{align}
      where $\rmtdeflate^2 = 1 + {\delta^2}c_1 / \rmtchisq$, $\rmtchisq =
      {\chi^2_{\no}}/ {\no}$, and $\chi^2_{\no}$ is the chi-squared distribution
      with $\no$ degrees of freedom.  Also, $\bar{\rho}_\nv > 0$ almost surely
      if $\gamma_{\beta, z}> 0$.  And the asymptotic improvement of the optimal
      angles $\theta_{\beta, \eb_{\rho^*}}$ and $\theta_{\beta,
      \eb_{\bar{\rho}_\nv}}$ over the original angle $\theta_{\beta, \eb}$ as
      $\nv\rightarrow \infty$ is,
      $$
        \frac{\sin^2\theta_{\beta, \eb_{\rho^*_{\nv}}}}
          {\sin^2\theta_{\beta, \eb}} \sas
          \frac{1 - \gamma_{\beta, z}^2}
            {1 - (\frac{\gamma_{\beta, z}}{\rmtdeflate})^2}
        = \frac{\sin^2\theta_{\beta, \eb_{\bar{\rho}_{\nv}}}}
            {\sin^2\theta_{\beta, \eb}}.
      $$


  \end{enumerate}

\end{thm}

\begin{remark}
  Geometrically, there are two views of $\eb_{\rho^*}$.  One is that
  $\eb_{\rho^*}$ is the projection of $\eb$ onto $\bbS_\beta$.  The
  other is that $\eb_{\rho^*}$ is the projection of $\beta$ onto the
  geodesic defined by $\eb_{\rho}$.  In either case, our goal is to find the
  intersection of the geodesic and the space $\bbS_\beta$.
\end{remark}

\begin{remark}
  While we consider a specific target, $z$, in principle the target does not matter.  It is possible that these kind of factor corrections can be
  applied beyond the first factor, given enough information to create a
  reasonable prior.
\end{remark}

The first takeaway from this result is that in the high
dimensional limit, it is always possible to  improve on the PCA
estimate by moving along the geodesic between $\hat\beta$  and
$z$.  As $\gamma_{\beta, z}$ approaches 0 or for a larger $\no$,
the optimal correction approaches 0. Conversely, as
$\gamma_{\beta, z}$ approaches 0 or for smaller $\no$, the
magnitude of the correction is larger. For $\gamma_{\beta, z} =
1$, the proper choice is naturally to choose $z$ since $\beta$
and $z$ are aligned in that case.  


The improvement in the angle as measured by the ratio of squared
sines is bounded in the interval $(1 - \gamma_{\beta, z}^2, 1)$.
As $\gamma_{\beta, z}$ approaches 0 or for larger $\no$, the
improvement diminishes and the ratio approaches 1.  Conversely,
for large values of $c_1$, the improvement approaches $1 -
\gamma_{\beta, z}^2$, indicating that improvement is naturally
constrained by how close $\beta$ is to $z$ in the first place.

In the application to the minimum variance portfolio, the
initial idea is to correct the sample eigenvector so that we
reduce the angle to the population eigenvector. However,  it is
not immediately clear that this should have a dramatic effect.
Even more surprising is that underestimation of risk has a large
component due to sample eigenvector bias and not any sample
eigenvalue bias. While an improved estimate $\eb_\rho$ has the
potential to greatly improve forecast risk, this represents only
a single dimension on which to evaluate the performance of a
portfolio.  We could be sacrificing small tracking error to the
true long-short minimum variance portfolio in exchange for
better forecasting.  That however is not the case here.

Since $\rmtchisq$ and $\rmtdeflate$ are unobservable 
non-degenerate random variables,
determining their realized values, even with asymptotic
estimates, is an impossible task.  Hence perfect corrections to
kill off the driving term of underestimation of risk are not
possible.  However, it is possible to make corrections that
materially improve risk forecasts.

\subsection{Eigenvalue corrections}

\input{sections/eigenvalue}

%% file: sections/shrink.tex

Given an estimate $\fee$, we analyze a parametrized correction 
of the form 
\begin{align}
  \fee\p{\rho} = \frac{\fee + \rho z}{\|\fee + \rho z\|} \pt .
\end{align}
for $\rho \in \bbR$. The curve $\fee\p{\rho}$ represents a
geodesic between $-z$ and $z$ that passes through $\fee$ as
$\rho$ passes the origin.  We select the optimal \tq{shrinkage}
parameter $\rho^*$ as the $\rho$ that minimizes the error in our
metrics $\te^2_{\minw}$ and $\scrR_{\minw}$ asymptotically.

\begin{figure}[htp]
\centering
\begin{subfigure}[t]{.45\textwidth}
\centering
\includegraphics[width=\linewidth]{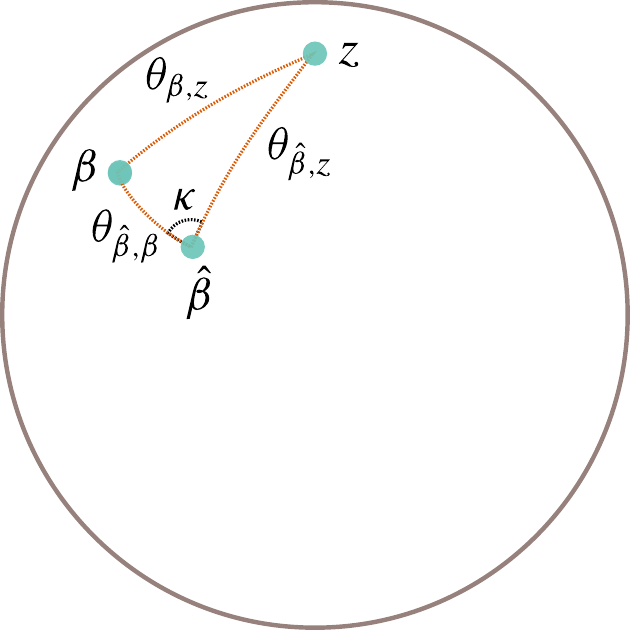}
 \caption{The spherical law of cosines states: \\
$\gamma_{\beta,z} - \gamma_{\eb,z}\gamma_{\eb,\beta}
= \sin \theta_{\eb,z} \sin \theta_{\eb,\beta} 
\cos \kappa$ where $\kappa$ is the angle of the corner
opposite $\theta_{\beta,z}$, arc-length between
$\beta$ and $z$.}\label{fig:cosines}
\end{subfigure}
\hspace{0.32in}
\begin{subfigure}[t]{.45\textwidth}
\centering
\includegraphics[width=\linewidth]{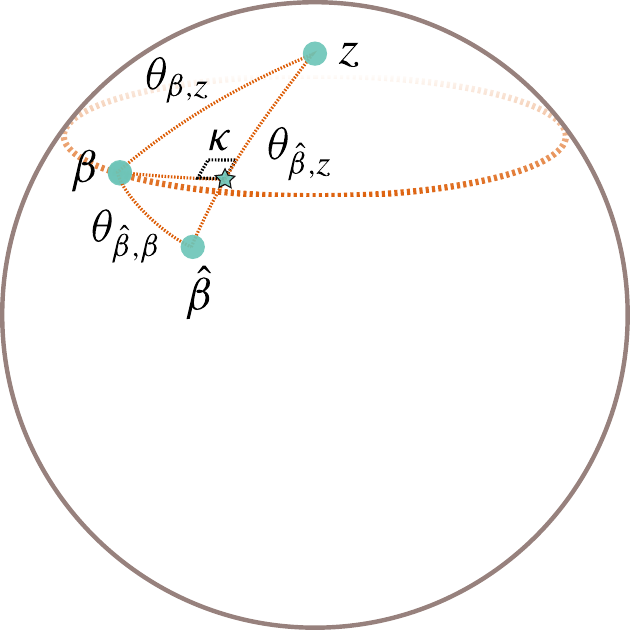}
 \caption{Setting the angle $\kappa = \pi/2$ corresponds 
to setting the error driving term 
$\err = 0$.
Thus, $\eb\p{\rho^*} \in \bbS_\beta$.}\label{fig:geodesic}
\end{subfigure}
\caption{Illustration of the spherical law of cosines and
the geodesic $\eb\p{\rho}$ between the points $\eb$ and $z$ 
along which we shrink the {\pca} estimate $\eb$. }
\label{F:cosines}
\end{figure}

With $\bbS^{\nv-1}$ denoting the unit $\nv$-sphere, 
we define the space 
\begin{align} \label{geodesic}
  \bbS_\fe &= \ff{ x \in \bbS^{\nv-1} 
  \cst  \gam{\fe}{z} - \gam{x}{z}\gam{x}{\fe} = 0}
   \pt .
\end{align}
\begin{lemma}\label{L:null}
Let $\fe \neq z$. There is a $\rho^*$ such that 
$\fee(\rho^*) \in \bbS_\fe$ and 
$\fee\p{\rho^*} \neq z$.
\end{lemma}
\begin{prove}
By the spherical law of cosines, we obtain
\begin{align}
   \gam{\fe}{z} - \gam{x}{z}\gam{x}{\fe}  = 
   \sin \ang{x}{z} \sin \ang{x}{\fe} \cos \kappa
\end{align}
where $\kappa$ is the angle (in $\bbS^{\nv-1}$)
 between the geodesic from 
$\fee$ to  $z$ and the one from $\fee$ to $\fe$ 
(see Figure \ref{fig:cosines}). Write $x = \fee\p{\rho}$
and $\kappa = \kappa_\rho$. Then, $\cos \kappa_\rho = 0$
when $\rho = \rho^*$ for which
the geodesic between $\fee\p{\rho^*}$ and $\fe$ is perperdicular
to that between $\fee$ and $z$.\footnote{If 
$\fee$ and $\fe$ do not lie on the same side of the 
sphere we must amend \eqref{geodesic} by replacing
$\fee$ with $-\fee$.  Note, $\gam{\fe}{z}
-\gam{\cdot}{z}\gam{\cdot}{\fe}$ has the same value for $x$
and $-x$.} By construction, $\fee(\rho^*)$ is not $z$ and 
is in $\bbS_\fe$.
\end{prove}

We observe that $\fee\p{\rho^*}$ of Lemma \ref{L:null} minimizes
the asymptotic error of our metrics. 
In particular, replacing $\fee$ with  $\fee\p{\rho^*}$ ensures 
that $\err$ is zero. Note that $\fee\p{\rho^*}
\neq z$ (unless $\fe = z$), i.e., we do not shrink $\fee$
all the way to $z$. While $z \in \bbS_\fe$, 
when $\beta \ne z$, the 
asymptotic error
$\err$ explodes as $\fee\p{\rho} \to z$. 

In a setting where $\fee$ is the {\pca} estimator, observe that
our selection of $\fee\p{\rho^*}$ does more than just correct
{\pca} bias. Theorem \ref{T:bias}, under the proper assumptions,
states that $\gam{\fe}{z} \sas \rmtdeflate \gam{\fee}{z}$ for a
random variable $\rmtdeflate > 1$. This suggests taking $\rho$
such that $\gam{\fe}{z}$ equals $\gam{\fee(\rho)}{z}$. This
choice is not optimal however as it lies on the contour $\f{ x
\in \bbS^{\nv-1} \cst \gam{x}{z} = \gam{\fe}{z}}$ (see Figure
\ref{fig:geodesic}).  It is not in $\bbS_\beta$ unless $\fe =z$
and its asymptotic error $\err$ is bounded away from zero.

%% file: sections/eigenvalue.tex
Our bias correction, based on formula \eqref{geodesic}, adjusts
the dominant eigenvector of the sample covariance matrix
$\mat{S}$.  It does not  involve standard corrections to the
eigenvalues of $\mat{S}$, which are well known to be biased.
This distinguishes our results from the existing literature (see
Section \ref{S:literature}).

For the purpose of improving accuracy of minimum variance
portfolios, there is no need to adjust the dominant eigenvalue.
As shown in formulas $\req{eq:gen-error}$ of
Section \ref{S:metrics},  the main drivers of $\te^2$ and
$\scrR$ for large minimum variance portfolios do not depend on
the dominant eigenvalue of $\mat{S}$ when returns follow a
one-factor model. This is a particular feature of minimum
variance, where the dominant factor is effectively hedged.


Since our correction removes a systematic form of bias, it can
be used to improve accuracy of other portfolios.  If these
portfolios have substantial factor exposure, however, a  {\it
compatibility adjustment} to the dominant eigenvalue may be
required. As our eigenvector adjustment, the compatibility
adjustment to the eigenvalue is distinct from the
eigenvalue corrections in the literature.



%

Here, we provide some discussion of compatible eigenvalue
corrections for \tq{simple} portfolios, i.e., those that do not
depend on the realized matrix of returns $\mat{R}$. Note, that
for these weights, the tracking error $\te$ is zero, so we treat
the risk forecast ratio $\scrR$ only.

Under assumptions on our one-factor model 
that hold for most cases of interest, one can write, for
a simple portfolio $w$ that
\begin{align} \label{eq:ratio_simple}
  \scrR_w \sim \frac{\vfe_{\nv}}{\vf_{\nv}} C \p{w, \beta,
\eb}^2
\end{align}
where $C\p{w,\beta,\eb} = (w^\top \beta)/(w^\top \eb)$. Our
correction addresses only the quantity $C$,  but asymptotic
formula
$\req{eq:ratio_simple}$ reveals that for simple portfolios,
sample eigenvalues play a material role. Another
difference between simple portfolios and minimum variance  is that the estimate $\vse$ of $\vs$ does not play
any role; the factor risk is all that matters. From the 
discusion in Section \ref{S:pca} we know that $\vfe_\nv$
tends to be larger than $\vf_{\nv}$ for large $\nv$, but it is not apriori clear
that an eigenvalue correction should aim to lower $\vfe_\nv$.
This would depend on the behavior of the coefficient
$C\p{w,\beta,\eb}$ given the estimate $\eb$ and the simple
portfolio $w$ at hand. Moreover, a correction that decreases
$\vfe_{\nv}$ will adjust risk forecast ratios downward,
potentially leading to unintended underforecasts. Thus,
sample eigenvalue corrections should be coupled with those 
for the sample eigenvectors to balance their respective
terms in $\req{eq:ratio_simple}$.

We state a sharp
result in this direction for the equally weighted portfolio,
a widely used simple portfolio.

\begin{proposition} \label{P:eqw}
Suppose Assumptions \ref{a:vol},
\ref{a:gauss}, \ref{a:orient}
hold. Let $\eqww = 1_N/N$ and define the corrected eigenvalue via
$$
\hat\sigma^2_{\rho^*_\nv} = 
   \Big( \frac{\gamma_{\eb, z}}{\gamma_{\eb_{\rho^*}, z}}
   \Big)^2 \vfe_N 
$$
where $\eb_{\rho^*}$ is our corrected PCA estimate $\eb$
of Theorem \ref{thm:main}. Then, the forecast variance ratio
satisfies $\scrR_{w} \sas \chi^2_T/T$ as $\nv \rightarrow
\infty$ where $\chi^2_T$ has a chi-squared distribution with
$T$ degrees of freedom.
\end{proposition}

Note that we adjust $\vfe_N$ downward since by design since
$\gamma_{\eb, z} \le \gamma_{\eb_{\rho^*}, z}$.

%% file: sections/algorithm.tex

For the precise statement of our algorithm see Appendix 
\ref{A:algorithm}. In what follows we address data-driven 
corrections for the case $\bdel = \vs \mat{I}$ of Theorem
\ref{thm:main} as well as the extension to heterogeneous
specific risk.

\subsection{Data-driven estimator for homogenous specific risk} 

We introduce procedure for constructing an estimator $\hat\rho$ for the
asymptotic oracle correction parameter $\bar{\rho}_\nv$ given in
\eqref{eq:rhostar_asy}.  It is based on estimates the specific variance
$\delta^2$ and $c_1$ from Assumption \ref{a:vol}.  From
\ci{yata2012}, we have the estimator given by,
\begin{align}\label{eq:c_hat}
  \hat{\delta}^2
    = \frac{\mathrm{Tr}(\mat{S}) - \hat\lambda_1}{\nv - 1 - \frac\nv\no},
\end{align}
where $\mat{S}$ is the sample covariance matrix for the data matrix $\mat{R}$
and $\hat\lambda_1 = \hat\sigma^2$ is the first eigenvalue of $\mat{S}$.  A
natural estimate for the true eigenvalue $\lambda_1(\bsig)$ is
$\wh{\lambda}_1^S = \max\{\hat\lambda_1 - \hat{\delta}^2 \nv / \no, 0\}$.
For $\hat\lambda_1$ sufficiently large, the estimate of $c_1$ is given by,
\begin{align}\label{eq:c1_hat}
  \hat{c}_1 
      = \frac{\nv}{ \no\hat{\lambda_1} - \nv\hat{\delta}^2 }.
\end{align}

Given the estimates of $\delta^2$ and $c_1$, we need a precise value for
$\rmtchisq$, as well as $\rmtdeflate$, in order to have a data-driven estimator.  We
approximate $\xi$ by its expectation $\mathbb{E}[\rmtchisq] = 1$ to obtain a
completely data driven correction parameter estimate $\hat\rho$,
\begin{align}\label{eq:rho_hat}
  \hat{\rho}
     = \frac{ \widehat{\rmtdeflate}\gamma_{\eb, z} }
          { 1 - (\widehat{\rmtdeflate}\gamma_{\eb, z})^2}
          \left(\widehat{\rmtdeflate} - \widehat{\rmtdeflate}^{-1}\right),
\end{align}
where $\widehat{\rmtdeflate}^2 = 1 + \hat{\delta}^2\hat{c}_1$.

We compute the factor variance as,
$$ 
  \hat\sigma^2_{\hat{\rho}} = \Phi_{\hat{\rho}}^2 \hat\sigma^2, 
    \quad \Phi_{\hat{\rho}} 
              = \frac{\gamma_{\eb, z}}{\gamma_{\eb_{\hat{\rho}}, z}},
$$
where $\hat\sigma^2 = \hat\lambda_1$ is the first sample eigenvalue from
$\mat{S}$.

\subsection{An extension to heterogenous specific risk} 
\label{S:heterogenous}
\input{sections/heterogenous}

%% file: sections/heterogenous.tex

Our analysis thus far has rested on the simplifying assumption that security
return specific variances have a common value.  Empirically, this is not the
case, and the numerical experiments discussed below in Section \ref{S:numerics}
allow for the more complex and realistic case of heterogenous specific
variances.   To address the issue, we modify both the oracle estimator and the
data-driven estimator by rescaling betas by specific variance. 

Under heterogeneous specific variance, the oracle value $\rho^*_\nv$ is given
by the formula,
\begin{align}\label{eq:rhostarhetero}
  \rho^*_\nv 
    = \frac{ 
          \gamma_{\beta, z}^{\wh{\bdel}} 
            - \gamma_{\beta, \eb}^{\wh{\bdel}} \gamma_{\eb, z}^{\wh{\bdel}}
        }{
          \gamma_{\beta, \eb}^{\wh{\bdel}} 
            - \gamma_{\beta, z}^{\wh{\bdel}} \gamma_{\eb, z}^{\wh{\bdel}}
        },
\end{align}
where $\gamma_{x, y}^{\wh{\bdel}} = x^T {\wh{\bdel}}^{-1} y / \sqrt{\gamma_{x,
x}^{\wh{\bdel}} \gamma_{y, y}^{\wh{\bdel}}} $ is a weighted inner product.
Furthermore, the risk adjusted returns $\mat{R}\bdel^{-1/2}$ have covariance
$\widetilde{\bsig}$ given by,
$$
  \widetilde{\bsig}
     = \vf \bdel^{-1/2}\beta \beta^\top \bdel^{-1/2} + \mat{I} \pt .
$$
For the risk adjusted returns, Theorem \ref{thm:main} holds.  The oracle formula
coupled with the risk adjusted returns suggest we use $\widetilde{\mat{R}} =
\mat{R}\wh{\bdel}^{-1/2}$ as the data matrix where $\wh{\bdel}$ is the specific
risk estimate from the standard PCA method.  Why should we expect this to work?
The purpose of the scaling $\mat{R}\bdel^{-1/2}$ is to make the specific return
distribution isotropic, and $\widetilde{\mat{R}}$ approximates that.  Since we
are only trying to obtain an estimator $\hat\rho$ that is close to $\rho^*_\nv$,
this approximation ends up fine.  And for ellipses specified by $\bdel$ with
relatively low eccentricity, the estimator in \eqref{eq:rho_hat}
actually works in practice since the distribution is relatively close to isotropic.  So for
larger eccentricity, we require the following adjustment just to get the data
closer to an isotropic specific return distribution.

The updated formulas for the heterogenous specific risk correction estimators
are given below, and we use them in our numerical experiments.  For an initial
estimate of specific risk $\wh{\bdel}$, the modified quantities are,
\begin{gather}
  \tilde{\eb} = \frac{ \wh{\bdel}^{-1/2}\eb }{ \|\wh{\bdel}^{-1/2}\eb\|_2 },
    \quad \tilde{z} = \frac{ \wh{\bdel}^{-1/2}z }{ \|\wh{\bdel}^{-1/2}z\|_2 },\\
  \widetilde{\mat{S}} = \wh{\bdel}^{-1/2} \mat{S} \wh{\bdel}^{-1/2},
    \quad \tilde{\hat\lambda}_1 = \hat\lambda_1 \|\wh{\bdel}^{-1/2}\eb\|^2,\\
  \tilde{\hat\delta}^2 
    = \frac{
        \mathrm{Tr}(\widetilde{\mat{S}}) - \tilde{\hat\lambda}_1
      }{ \nv - 1 - \frac{\nv}{\no} },
  \quad
  \hat{c}_1 
    = \frac{\nv}{\no\tilde{\hat\lambda}_1 - \nv\tilde{\hat\delta}^2 },\\
  \hat{\rho}
     = \frac{ \widehat{\rmtdeflate}\gamma_{\tilde{\eb}, \tilde{z}} }
          { 1 - (\widehat{\rmtdeflate}\gamma_{\tilde{\eb}, \tilde{z}})^2}
          \left(\widehat{\rmtdeflate} - \widehat{\rmtdeflate}^{-1}\right).
\end{gather}
We use the PCA estimate of the specific risk $\wh{\bdel} = \text{\bf diag}\p{
\mat{S} - \hat\sigma^2 \eb \eb^\top}$ as the initial estimator.

Once we have the estimated $\hat\rho$, we return to the original data matrix
$\mat{R}$ and apply the correction as before to $\hat\beta$, the first
eigenvector of the sample covariance matrix.  The method for correcting the
sample eigenvalue remains the same and we opt to recompute the specific
variances using the corrected factor exposures and variance.  


%% file: sections/numerics.tex

We use simulation to quantify the dispersion bias  and  its
mitigation in minimum variance and equally portfolios.  
We design our simulations around a return generating process that is  more
realistic than the single-factor, homogenous-specific-risk model featured in
Theorems~\ref{thm:pca_bias} and ~\ref{thm:main}.\footnote{The simplistic setting
for our theoretical results showcases the main theoretical tools used in their
proofs without the distraction of regularity conditions required for
generalization. Global equity risk models used by investors typically include
more than 100 factors. }   Returns to  $\nv \in \bbN$ securities are generated
by a multi-factor model with heterogenous specific risk,
\begin{align} 
  R =  \phi B+ \ep, \label{multifactor}
\end{align}
where $\phi$ is the vector of factor returns,  $B$ 
is the matrix of factor exposures, and $\ep = \p{\ep^1, \ep^2, 
\dots, \ep^{\nv}}$ is the vector of diversifiable specific  returns. 

Our examples are based on a model with $K = 4$ factors.  For consistency with
Sections~\ref{S:problem} and~\ref{S:main}, we  continue to adopt the
mathematically convenient convention of scaling exposure vectors to have $L_2$
norm 1, and we denote the vector of exposures to the first factor (the first
column of the exposure matrix $B$) by $\beta$.   The recipe for constructing
$\beta$ with a target value $\gamma_{\beta, z}$ is to generate a random vector,
rescale the vector to length 1, and then modify the component in the $z$
direction to have the correct magnitude (while maintaining length 1).  A similar
approach would be to construct a random vector $\beta'$ with aveage equal to 1
and variance equal to $\tau^2$, where $\tau^2$ is the dispersion of the ``market
beta,'' and then rescale $\beta'$ to length 1 to obtain $\beta$.  The parameter
$\tau^2$ would control the concentration of the market betas, just like
$\gamma_{\beta, z}$, and tends to be greater in calmer regimes.  The connection
between $\tau^2$ and the dispersion parameter $\gamma_{\beta, z}$ is given by
$\gamma_{\beta, z}^2 = 1 / \sqrt{ 1 + \tau^2 }.$



 The three remaining factors are fashioned from equity styles such as
 volatility, earnings yield and size.\footnote{Seminal references on the
 importance of  volatility (or beta), earnings yield and  size in explaining
 cross-sectional correlations is in \ci{black1972}, \ci{basu1983} and
 \ci{banz1981}.}  We draw the exposure of each security to each factor from a
 mean 0, variance 0.75 normal distribution and again normalize each vector of
 factor exposures to have $L_2$ length 1. 
 
We calibrate the risk of the market factor in accordance with  \ci{clarke2011} and \ci{goldberg2014}; both report the annualized volatility of the US market to be roughly 16\%, based on estimates that rely on decades of data.    We calibrate the risk of factors 2, 3 and 4 in accordance with  \ci[Table 4.3]{morozov2012}, by setting their annualized volatilities to be 8\%, 4\% and 4\%. 
We assume that the returns to the factors are pairwise uncorrelated.  

We draw annualized specific volatilities 
$\f{\vs_n}$ from a uniform distribution on 
$[32\%, 64\%]$.  This range is somewhat broader than the
estimated range in \ci{clarke2011}.\footnote{For example,
the empirically
observed range in
\ci{clarke2011} is  $[25\%, 65\%]$.}

In each experiment, we  simulate a year's worth of daily
returns,  $\no = 250$, to $\nv$ securities.  From this data set,
we construct a sample covariance matrix, $\mat{S}$,  from which
we extract three estimators of the factor covariance matrix
$\bsig$.  The first is the data-driven estimator, the
implementation specifics of which are discussed in Section
\ref{S:heterogenous} and precisely summarized of Appendix
\ref{A:algorithm}. The second is the \emph{oracle} estimator
which is the same as the data-driven estimator but with the
true value of $\gamma_{\beta,z}$, the projection of the true
factor onto the $z$-vector, supplied.\footnote{For the data-driven
estimator, this quantity is estimated from the observed data.}
Our third estimator is classical PCA, which is specified in
detail in Section \ref{S:pca}.  We use the three estimated
covariance matrices to construct minimum variance portfolios and
to forecast their risk. We also use these covariance matrices to
forecast the risk of an equally weighted portfolio.

\begin{figure}[H]
\includegraphics[width=\textwidth]{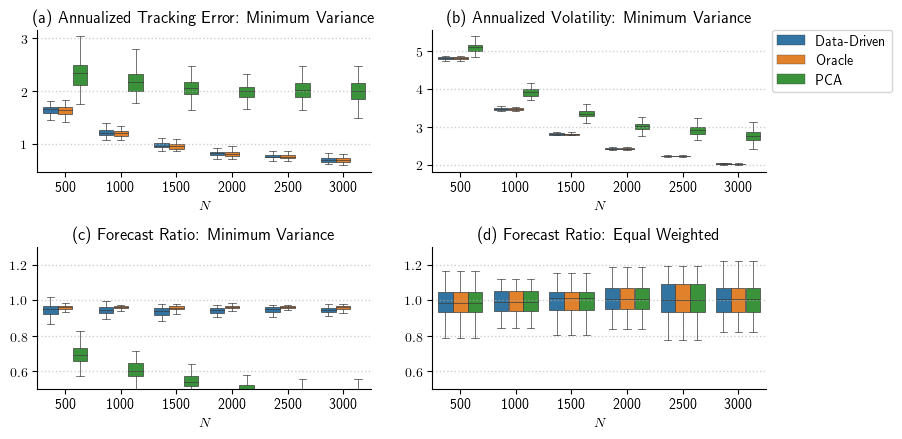}
\caption{Performance statistics for 50 simulated data sets of $\no= 250$
observations as $\nv$ varies and $\gamma_{\beta, z} = 0.9$.  Panel (a):
Annualized tracking error for minimum variance portfolios.  Panel (b):
Annualized volatility for minimum variance portfolios.  Panel (c):  Variance
forecast ratio for minimum variance portfolios.  Panel (d): Variance forecast
ratio for equally weighted portfolios. \label{figure:varyN}}
\end{figure}

\begin{figure}[H]
\includegraphics[width=\textwidth]{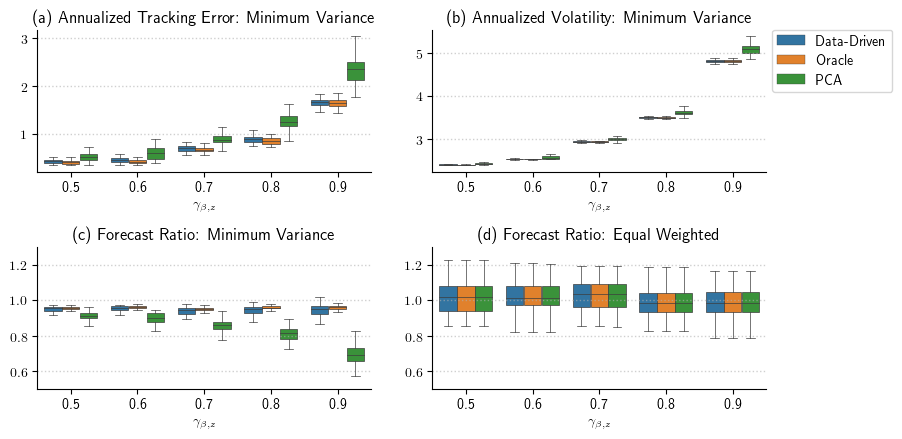}
\caption{Performance statistics for 50 simulated data sets of $\no= 250$
observations  as $\gamma_{\beta, z}$ varies and $\nv = 500$.  Panel (a):
Annualized tracking error for minimum variance portfolios.  Panel (b):
Annualized volatility for minimum variance portfolios.  Panel (c):  Variance
forecast ratio for minimum variance portfolios.  Panel (d): Variance forecast
ratio for equally weighted portfolios.\label{figure:varydisp}}
\end{figure}

\begin{figure}[H]
\includegraphics[width=\textwidth]{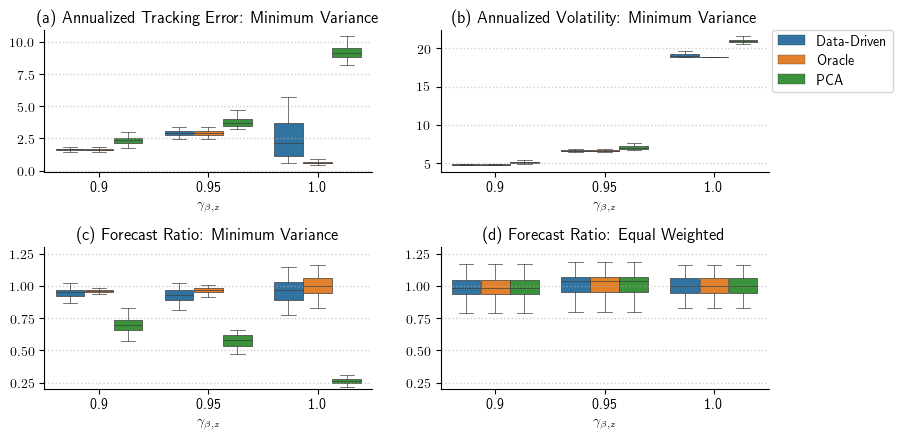}
\caption{Performance statistics for 50 simulated data sets of $\no= 250$
observations  as $\gamma_{\beta, z}$ varies and $\nv = 500$.  Panel (a):
Annualized tracking error for minimum variance portfolios.  Panel (b):
Annualized volatility for minimum variance portfolios.  Panel (c):  Variance
forecast ratio for minimum variance portfolios.  Panel (d): Variance forecast
ratio for equally weighted portfolios.\label{figure:varydispconc}}
\end{figure}

In the experiments described below, we vary the number of securities, $\nv$, and the concentration of the market factor, $\gamma_{\beta, z}$. We run 50 simulations for each pair,  $(\nv, \gamma_{\beta, z})$. Results are shown in Figure~\ref{figure:varyN} for 
a fixed concentration and varying number of securities, and in
Figures~\ref{figure:varydisp} and~\ref{figure:varydispconc} for a fixed number of securities and varying concentration.  
Panels (a) and (b) in each figure show 
annualized tracking error  and volatility for a minimum variance portfolio. 
Panels (c) and (d) show variance forecast ratio for a
minimum variance portfolio and an equally
weighted portfolio.

In Figure~\ref{figure:varyN},  the concentration of the market factor is $\gamma_{\beta, z} = 0.9$.  Panels (a), (b) and (c) show that for minimum variance portfolios optimized with dispersion bias-corrected PCA models, tracking error and volatility decline materially as $\nv$ grows from 500 to 3000,  ranges of outcomes compress, and variance forecast ratios are near 1 for all $\nv$ considered. These desirable effects are less pronounced, or even absent, in a PCA model without dispersion bias mitigation.  A comparison of  panels (c) and (d) highlights the difference in accuracy of risk forecasts between a minimum variance portfolio and an equally weighted portfolio.  Dispersion bias mitigation materially improves variance forecast ratio for the former and has no discernible impact for the latter.  


In Figure~\ref{figure:varydisp}, the number of securities is $\nv = 500$.  Panels (a), (b) and (c) show that for minimum variance portfolios optimized  with PCA models, tracking error and volatility increase materially as $\gamma_{\beta, z}$ grows from 0.5 to 0.9, variance forecast ratio diminishes,  and the ranges of outcomes expand. These undesirable effects are diminished when the dispersion bias is mitigated.  

Results for values $\gamma_{\beta, z} \in [0.9, 1.0]$ are shown separately in Figure~\ref{figure:varydispconc}. The severe decline in performance for all risk metrics for   $\gamma_{\beta, z}$ in this range is a consequence of the sea change in the composition of a minimum variance portfolio that can occur when the dominant eigenvalue of the covariance matrix is sufficiently concentrated.     Here, the true minimum variance portfolio tends to lose its short positions.\footnote{If specific variances are all equal, the minimum variance portfolio becomes equally weighted when the dominant eigenvector is dispersionless.}  Errors in estimation of the dominant factor lead to long-short optimized portfolios approximating long-only optimal portfolios.   The market factor is hedged in the former but not in the latter, and this discrepancy propagates to the error metrics.

A comparison of  panels (c) and (d)  in Figures~\ref{figure:varydisp} and~\ref{figure:varydispconc} highlights, again,  the difference in accuracy of risk forecasts between a minimum variance portfolio and an equally weighted portfolio.  Dispersion bias mitigation materially improves variance forecast ratio for the former and has no discernible impact on the latter.

A casual inspection of the figures suggests that the data-driven estimator performs
nearly as well as the oracle.  However, there can be
substantial differences between the two estimators in some cases, as shown in 
tracking error and variance forecast ratio of minimum variance
Figure~\ref{figure:varydispconc}  when $\gamma_{\beta, z} \in
[0.9, 1.0]$.  The origin of the differences can be seen
Lemma~\ref{L:null}.  In order for the tracking error
and variance forecast ratio to have good asymptotic properties,
the estimator $\eb$ must lie in the null space
$\bbS_\beta$.  This  condition is guaranteed for the
oracle estimators but will not generally be satisfied for
data-driven estimators.

%% file: sections/summary.tex

 In this article,  we develop a correction for bias in  PCA-based covariance matrix estimators.  The bias is excess dispersion in a dominant eigenvector, and the form of the correction is suggested by formulas for estimation error metrics applied to minimum variance portfolios.

We identify an oracle correction that optimally shrinks the sample eigenvector along a spherical geodesic toward the distinguished zero-dispersion vector, and we provide asymptotic guarantees that oracle shrinkage reduces both types of error.  These findings  are especially relevant to equity return covariance matrices,  which feature a dominant  factor whose overwhelmingly positive exposures tend to be overly dispersed by PCA estimators.  

Our results fit into two streams of academic literature.  The first is the large-$N$-fixed-$T$ branch of random matrix theory, which belongs to statistics.  The second is empirical finance, which features results about Bayesian adjustments to estimated betas.

To enable practitioners to use our results, we develop a data-driven estimator of the oracle.  Simulation experiments support the practical value of our correction, but much work remains to be done.  That includes the development of estimates of the size and likelihood of the exposure bias in finite samples, the identification and correction of biases in other risk factors,  and empirical studies.  Explicit formulas for error metrics in combination with the geometric perspective in this article provide a way to potentially improve construction and risk forecasts of investable optimized portfolios.

%% file: appendix/algorithm.tex

Our corrected covariance matrix algorithm is given in Algorithm
\ref{alg:eigenvector_correction} where the input is a data matrix of returns
$R$.

\begin{algorithm}
  \caption{1-Factor Bias Corrected PCA Covariance Estimator}
  \label{alg:eigenvector_correction}
  \begin{algorithmic}[1]
    \Require $\mat{R} = \p{R_1, \dots, R_{\no}}$
    \Require $z = 1_{\nv} / \sqrt{\nv}$
    \Procedure{Bias Corrected PCA Covariance}{$\mat{R}$}

      \State $\mat{S} \gets \frac{1}{\no} \mat{R} \mat{R}^T$
      \State 
          $\hat{\mat{U}} \gets [\hat{u}_1, \ldots, \hat{u}_\nv]$,
          \quad $\hat{\mat{\Lambda}} 
            \gets \mathrm{Diag}(\hat{\lambda}_1, \ldots, \hat{\lambda}_N)$
          \Comment{Eigendecomposition of $\mat{S}$}

      \State $\eb \gets \mathrm{sign}(\hat{u}_1^T z)\hat{u}_1$
        \Comment{Orient such that $\eb^T z > 0$.}
      \State $\hat\sigma^2 \gets \hat{\lambda}_1$

      \State $\wh{\bdel} \gets \mathrm{Diag}(\mat{S} - \wh{\bell})$,
             \quad
             $\wh{\bell} \gets \hat\sigma^2 \eb\eb^T$
             \Comment{Initial PCA estimate.}
      
      \State $\widetilde{\mat{S}} \gets
                \wh{\bdel}^{-1/2} \mat{S} \wh{\bdel}^{-1/2}$,
             \quad
             $\tilde{\hat\lambda}_1 \gets 
                 \hat\lambda_1 \|\wh{\bdel}^{-1/2}\eb\|^2$
      \Statex $\tilde{\eb} \gets
                \frac{ \wh{\bdel}^{-1/2}\eb }{ \|\wh{\bdel}^{-1/2}\eb\|_2 }$,
             \quad
             $\tilde{z} \gets
                 \frac{ \wh{\bdel}^{-1/2}z }{ \|\wh{\bdel}^{-1/2}z\|_2 }$
      \State $\widehat{\rmtdeflate} \gets 1 + \tilde{\hat\delta}^2 \hat{c}_1$,
             \quad
             $\tilde{\hat\delta}^2  \gets
                \frac{
                  \mathrm{Tr}(\widetilde{\mat{S}}) - \tilde{\hat\lambda}_1
                }{ \nv - 1 - \frac{\nv}{\no} }$,
             \quad
             $\hat{c}_1 \gets
              \frac{\nv}{\no\tilde{\hat\lambda}_1 - \nv\tilde{\hat\delta}^2}$
      \State $\eb_{\hat{\rho}} \gets 
                \frac{\eb + \hat{\rho}z}
                  {\sqrt{1 + 2\hat{\rho}\gamma_{\eb, z} + \hat{\rho}^2}}$,
             \quad
             $\hat{\rho} \gets
                \frac{ \widehat{\rmtdeflate}\gamma_{\tilde{\eb}, \tilde{z}} }
                    { 1 - (\widehat{\rmtdeflate}\gamma_{\tilde{\eb}, \tilde{z}})^2}
                \left(\widehat{\rmtdeflate} - \widehat{\rmtdeflate}^{-1}\right)$

      \State $\wh{\bdel}_{\hat\rho} \tabularnewline\tabularnewline
                \gets \mathrm{Diag}(\mat{S} - \wh{\bell}_{\hat\rho})$,
             \quad
             $\wh{\bell}_{\hat\rho} 
                \gets \hat\sigma^2_{\hat\rho} \eb_{\hat{\rho}}\eb_{\hat{\rho}}^T$,
             \quad
             $\hat\sigma^2_{\hat\rho}
                \gets \frac{\gamma_{\eb, z}^2}{\gamma_{\eb_{\hat{\rho}}, z}^2} 
                          \hat\sigma^2$
      \State \textbf{return} 
          $\hat{\bsig} \gets \wh{\bell}_{\hat\rho} + \wh{\bdel}_{\hat\rho}$ 
          \Comment{The 1-factor bias corrected PCA estimator.}
    \EndProcedure
  \end{algorithmic}
\end{algorithm}






%% file: tables/tables.tex
\input{tables/table_N.tex}

\input{tables/table_gam.tex}

%% file: tables/table_N.tex
\begin{table}[htp]
  \centering
  \input{tables/multifactor_highrisk/table_N.tex}

  \caption{Median values derived from Figure \ref{figure:varyN}. Performance
    statistics for 50 simulated data sets of $\no= 250$ observations as
    $\gamma_{\beta, z}$ varies and $\nv = 500$.  Panel (a): Annualized tracking
    error for minimum variance portfolios.  Panel (b): Annualized volatility for
    minimum variance portfolios.  Panel (c):  Variance forecast ratio for
    minimum variance portfolios.  Panel (d): Variance forecast ratio for equally
    weighted portfolios. \label{table:varyN}
  }

\end{table}

%% file: tables/multifactor_highrisk/table_N.tex
\begin{subtable}[t]{\linewidth}
    \centering
    \begin{tabular}{lrrrrrr}
\toprule
N &  500  &  1000 &  1500 &  2000 &  2500 &  3000 \\
\midrule
PCA         & 2.35\% & 2.18\% & 2.06\% & 2.01\% & 2.03\% & 2.00\% \\
Oracle      & 1.64\% & 1.21\% & 0.97\% & 0.83\% & 0.77\% & 0.70\% \\
Data-Driven & 1.66\% & 1.21\% & 0.98\% & 0.84\% & 0.77\% & 0.71\% \\
\bottomrule
\end{tabular}

    \caption{Annualized Tracking Error: Minimum Variance}
\end{subtable}

\begin{subtable}[t]{\linewidth}
    \centering
    \begin{tabular}{lrrrrrr}
\toprule
N &  500  &  1000 &  1500 &  2000 &  2500 &  3000 \\
\midrule
PCA         & 5.10\% & 3.92\% & 3.34\% & 3.03\% & 2.91\% & 2.75\% \\
Oracle      & 4.82\% & 3.47\% & 2.80\% & 2.42\% & 2.22\% & 2.01\% \\
Data-Driven & 4.82\% & 3.47\% & 2.81\% & 2.42\% & 2.23\% & 2.02\% \\
\bottomrule
\end{tabular}

    \caption{Annualized Volatility: Minimum Variance}
\end{subtable}

\begin{subtable}[t]{\linewidth}
    \centering
    \begin{tabular}{lrrrrrr}
\toprule
N &  500  &  1000 &  1500 &  2000 &  2500 &  3000 \\
\midrule
PCA         & 0.695 & 0.604 & 0.541 & 0.492 & 0.452 & 0.413 \\
Oracle      & 0.963 & 0.962 & 0.960 & 0.961 & 0.963 & 0.961 \\
Data-Driven & 0.951 & 0.946 & 0.943 & 0.945 & 0.953 & 0.947 \\
\bottomrule
\end{tabular}

    \caption{Forecast Ratio: Minimum Variance}
\end{subtable}

\begin{subtable}[t]{\linewidth}
    \centering
    \begin{tabular}{lrrrrrr}
\toprule
N &  500  &  1000 &  1500 &  2000 &  2500 &  3000 \\
\midrule
PCA         & 0.983 & 0.991 & 1.014 & 1.011 & 1.001 & 1.006 \\
Oracle      & 0.983 & 0.991 & 1.015 & 1.011 & 1.001 & 1.006 \\
Data-Driven & 0.983 & 0.991 & 1.015 & 1.011 & 1.001 & 1.006 \\
\bottomrule
\end{tabular}

    \caption{Forecast Ratio: Equal Weighted}
\end{subtable}

%% file: tables/table_gam.tex
\begin{table}[htp]
  \centering
  \input{tables/multifactor_highrisk/table_gam.tex}

  \caption{Median values derived from Figures \ref{figure:varydisp} and
    \ref{figure:varydispconc}.  Performance statistics for 50 simulated data
    sets of $\no= 250$ observations  as $\gamma_{\beta, z}$ varies and $\nv =
    500$.  Panel (a): Annualized tracking error for minimum variance portfolios.
    Panel (b): Annualized volatility for minimum variance portfolios.  Panel
    (c):  Variance forecast ratio for minimum variance portfolios.  Panel (d):
    Variance forecast ratio for equally weighted
    portfolios.\label{table:varydisp} }
    
\end{table}

%% file: tables/multifactor_highrisk/table_gam.tex
\begin{subtable}[t]{\linewidth}
    \centering
    \begin{tabular}{lrrrrrrr}
\toprule
$\gamma_{\beta, z}$ &  0.50 &  0.60 &  0.70 &  0.80 &  0.90 &  0.95 &  1.00 \\
\midrule
PCA         & 0.51\% & 0.59\% & 0.88\% & 1.26\% & 2.35\% & 3.72\% & 9.11\% \\
Oracle      & 0.41\% & 0.42\% & 0.67\% & 0.86\% & 1.64\% & 2.91\% & 0.63\% \\
Data-Driven & 0.43\% & 0.45\% & 0.70\% & 0.89\% & 1.66\% & 2.94\% & 2.18\% \\
\bottomrule
\end{tabular}

    \caption{Annualized Tracking Error: Minimum Variance}
\end{subtable}

\begin{subtable}[t]{\linewidth}
    \centering
    \begin{tabular}{lrrrrrrr}
\toprule
$\gamma_{\beta, z}$ &  0.50 &  0.60 &  0.70 &  0.80 &  0.90 &  0.95 &   1.00 \\
\midrule
PCA         & 2.42\% & 2.57\% & 2.99\% & 3.62\% & 5.10\% & 7.02\% & 20.95\% \\
Oracle      & 2.40\% & 2.54\% & 2.94\% & 3.50\% & 4.82\% & 6.63\% & 18.88\% \\
Data-Driven & 2.41\% & 2.54\% & 2.95\% & 3.51\% & 4.82\% & 6.65\% & 18.99\% \\
\bottomrule
\end{tabular}

    \caption{Annualized Volatility: Minimum Variance}
\end{subtable}

\begin{subtable}[t]{\linewidth}
    \centering
    \begin{tabular}{lrrrrrrr}
\toprule
$\gamma_{\beta, z}$ &  0.50 &  0.60 &  0.70 &  0.80 &  0.90 &  0.95 &  1.00 \\
\midrule
PCA         & 0.915 & 0.904 & 0.860 & 0.816 & 0.695 & 0.579 & 0.264 \\
Oracle      & 0.958 & 0.965 & 0.949 & 0.959 & 0.963 & 0.967 & 1.001 \\
Data-Driven & 0.955 & 0.959 & 0.945 & 0.949 & 0.951 & 0.928 & 0.967 \\
\bottomrule
\end{tabular}

    \caption{Forecast Ratio: Minimum Variance}
\end{subtable}

\begin{subtable}[t]{\linewidth}
    \centering
    \begin{tabular}{lrrrrrrr}
\toprule
$\gamma_{\beta, z}$ &  0.50 &  0.60 &  0.70 &  0.80 &  0.90 &  0.95 &  1.00 \\
\midrule
PCA         & 1.019 & 1.015 & 1.035 & 0.984 & 0.983 &  1.04 & 1.002 \\
Oracle      & 1.020 & 1.015 & 1.035 & 0.984 & 0.983 &  1.04 & 1.002 \\
Data-Driven & 1.020 & 1.015 & 1.035 & 0.984 & 0.983 &  1.04 & 1.002 \\
\bottomrule
\end{tabular}

    \caption{Forecast Ratio: Equal Weighted}
\end{subtable}

%% file: appendix/proofs.tex

\newcommand{\sph}[1]{B(#1)}

We start off with some foundational asymptotic results from the literature.  Let
$\mat{X} \sim \mathcal{N}(0, \blam)$ where $\blam = \text{\bf diag}(\lambda_1,
\lambda_2, \ldots, \lambda_2)$ is a diagonal matrix satisfying with $\lambda_1$
satisying Assumption $\mathcal{C}1$ in \ci{shen2016}.  Let $\mat{S}_\mat{X} =
\frac{1}{\no} \mat{X}\mat{X}^T$ be the sample covariance for $\mat{X}$ with
eigendecomposition $\mat{S}_\mat{X} = \hat{\mat{V}} \hat{\blam}
\hat{\mat{V}}^T$.  Further let $\hat{v}_1$ be the first sample eigenvector given
by,
$$
  \hat{v}_1
    = \begin{bmatrix} 
        \hat{v}_{11} \\ \ldots \\ \hat{v}_{\nv1} 
    \end{bmatrix},
$$
and define $\tilde v = \begin{bmatrix} \hat{v}_{21} & \ldots &
\hat{v}_{\nv1} \end{bmatrix}^T$.  By \ci[Theorem 6]{paul2007}, $\tilde v
\sim \mathrm{Unif}(\sph{\nv - 2})$ where $\sph{\nv - 2}$ is a unit
$\nv - 2$ sphere. 

Via a simple scaling by $\lambda_2$, by \ci[Theorem
6.3]{shen2016} we have $e_1^T \hat{v}_1 = \hat{v}_{11} \cas
\rmtdeflate^{-1}$ where $\rmtdeflate^2 = 1 + \lambda_2 c_1 /
\rmtchisq$ and $\rmtchisq = {\chi_{\no}^2} / \no$.

We introduce the following lemma, which we will use in the
proofs of our results. We leave its proofs until the end.

\begin{lemma}\label{L:main}

  If $X_n \sim \mathrm{Unif}(\sph{n - 1})$ and $Y_n \in \sph{n - 1}$ is a
  sequence independent of $X_n$, then $X_n^T Y_n \cas 0$ as $n \rightarrow
  \infty$.


\end{lemma}

\begin{proof}[\textsc{Proof of Theorem \ref{T:bias}}]  

  Let $\mat{X} = \mat{U} \mat{R}$ where $\mat{U}$ is the matrix of eigenvectors
  $(\beta, u_2, \ldots u_{\nv})$ of $\bsig$ so that $\mathrm{Cov}(\mat{X}) =
  \blam$ as introduced in the beginning of this section.  Also as before let
  $\mat{S}_\mat{X} = \frac{1}{\no} \mat{X}\mat{X}^T = \hat{\mat{V}}
  \hat{\blam} \hat{\mat{V}}^T$ be the sample covariance of $\mat{X}$ and its
  eigendecomposition.  Then $\mat{S} = \frac{1}{\no} \mat{R}\mat{R}^T =
  \mat{U}\hat{\mat{V}} \hat{\blam} \hat{\mat{V}}^T\mat{U}^T$ so that the first
  sample eigenvector of $\mat{S}$, $\eb$, is given by,
  $$
    \eb = \hat{v}_{11} \beta + \sum_{j=2}^\nv \hat{v}_{j1}u_j.
  $$
  Then we have $\gamma_{\eb, \beta} = \hat{v}_{11}$ and,
  $$
    \gamma_{\eb, z}
       = \hat{v}_{11} \gamma_{\beta, z} + \tilde{v}^T \omega_{\nv},
  $$
  where $\omega_N = \begin{bmatrix} u_2^T z & \cdots & u_{\nv}^T z
  \end{bmatrix}$.  As noted before, by \ci[Theorem
6.3]{shen2016}, 
  $\hat{v}_{11} \pt \cas \pt \rmtdeflate^{-1}$.  We know that both
  $\|\tilde{v}\|_2$ and $\|\omega_{\nv}\|_2$ are bounded as,
  $$
    \|\tilde{v}\|_2 \leq 1, \quad 1 = \|U^Tz\|^2_2 = (\beta^Tz)^2 + \sum (u_j^T
    z)^2 = \gamma_{\beta, z}^2 + \|\omega_N\|^2_2.
  $$
  Therefore, from Lemma \ref{L:main} for $X_\nv =
  \frac{\tilde{v}}{\|\tilde{v}\|_2}$ and $Y_\nv =
  \frac{\omega_{\nv}}{\|\omega_{\nv}\|_2}$ we have $\tilde{v}^T \omega_{\nv}$
  converges almost surely to 0 so we conclude the result.


\end{proof}

\begin{proof}[\textsc{Proof of Theorem \ref{thm:main}}] 
  ~
  \begin{enumerate}[i.]

    \item It is easy to verify that $\rho^*_{\nv}$ solves 
  $\gam{\fe}{z} - \gam{\fee(\rho)}{z}\gam{\fee(\rho)}{\fe}$ 
  and maximizes $\gamma_{\beta, \eb_{\rho}}$.

    \item For the oracle value $\rho^*_{\nv}$, 
     $\eb_{\rho^*_{\nv}} \in \mathcal{S}_\beta$ so the result 
     follows by Lemma~\ref{L:null}.

    \item  Convergence of $\rho^*_{\nv}$ to $\bar{\rho}_{\nv}$ 
stems from Theorem\ref{T:bias}.  It is also clear that $\bar{\rho}_{\nv}$ if
      $\gamma_{\beta, z} > 0$ since $\rmtdeflate - \rmtdeflate^{-1} > 0$.

      For the asymptotic improvement due to shrinkage, we rely on \ci[Theorem
      1]{bjorck1973}, which shows that principal angle can be derived from the
      singular value decomposition of,
      $$
        \beta^T \begin{bmatrix} 
          z & 
          \frac{\eb - \gamma_{\eb, z}z}{\sqrt{1 - \gamma_{\eb, z}^2}}
        \end{bmatrix} = \begin{bmatrix} 
          \gamma_{\beta, z} & 
          \frac{\gamma_{\beta, \eb} - \gamma_{\eb, z}\gamma_{\beta, z}}
              {\sqrt{1 - \gamma_{\eb, z}^2}}
        \end{bmatrix}.
      $$
      By maximizing $\gamma_{\beta, \eb_\rho}$ (or equivalently minimizing the
      angle between $\beta$ and $\beta_\rho$), we are directly choosing
      $\rho^*_{\nv}$ such that $\eb_{\rho^*_{\nv}}$ is the principal vector with
      corresponding principal angle to $\eb$.  Finding the vector in terms of
      correction quantity is easier through direct maximization of
      $\gamma_{\beta, \eb_\rho}$, and finding the improvement is easier through
      the principal angle computation, despite the results being equivalent.

      From the above product, the squared cosine of the principal angle and its
      asymptotic value is,
      \begin{align*}
        \gamma_{\beta, \eb_{\rho^*_{\nv}}}^2 
          & = \left\| 
              \begin{bmatrix} 
                \gamma_{\beta, z} & 
                \frac{\gamma_{\beta, \eb} - \gamma_{\eb, z}\gamma_{\beta, z}}
                    {\sqrt{1 - \gamma_{\eb, z}^2}}
              \end{bmatrix}
            \right\|_2^2 \\
          & = \gamma_{\beta, z}^2 
            + \frac{(\gamma_{\beta, \eb} - \gamma_{\eb, z}\gamma_{\beta, z})^2}
                {1 - \gamma_{\eb, z}^2} \\
          & \sas \gamma_{\beta, z}^2 
              + \frac{\rmtdeflate^{-2}(1 - \gamma_{\beta, z}^2)^2}
                    {1 - (\frac{\gamma_{\beta, z}}{\rmtdeflate})^2} \\
          & = \gamma_{\beta, \eb_{\bar{\rho}_{\nv}}}^2
      \end{align*}
      Since $\sin^2\theta_{\beta, \eb} = 1 - \gamma_{\beta, \eb}^2 \cas 1 -
      \rmtdeflate^{-2}$, we see clearly that,
      $$
        \frac{\sin^2\theta_{\beta, \eb_{\rho^*_{\nv}}}}
          {\sin^2\theta_{\beta, \eb}} \sas
          \frac{1 - \gamma_{\beta, z}^2}
            {1 - (\frac{\gamma_{\beta, z}}{\rmtdeflate})^2}
        = \frac{\sin^2\theta_{\beta, \eb_{\bar{\rho}_{\nv}}}}
            {\sin^2\theta_{\beta, \eb}}
      $$

  \end{enumerate}

\end{proof}

\begin{proof}[\textsc{Proof of Lemma \ref{L:main}}]

  By orthogonal invariance of $X_n$ and the independence of $Y_n$,
  $$
    X_n^T Y_n 
      = X_{n}^T Q_{n}^T Q_{n} Y_n 
      = (Q_n X_n)_1 
      \stackrel{\mathcal{D}}{=} X_{n1}
  $$
  where $Q_n$ is an orthongal matrix such that $Q_n Y_n = e_1$, $e_1$ is
  the first canonical vector, and $X_{n1}$ is the first entry of $X_{n}$.
  We know from \ci{muller1959} for $X_{n1}$,
  $$
    X_{n1} 
      \stackrel{\mathcal{D}}{=} \frac{Z_1}{\sqrt{Z_1^2 + \chi_{n - 1}^2}},
  $$
  where $Z_1 \sim \mathcal{N}(0, 1)$ and is independent of $\chi_{n - 1}^2$.
  We have,
  \begin{align*}
    \mathbb{P}(|X_{n1}| \geq \epsilon ) 
      \leq \frac{\mathbb{E}X_{n1}^4}{\epsilon^4}
      \leq \frac{1}{\epsilon^4n^{2}} 
          \mathbb{E}\left[Z_1^4\right] 
          \mathbb{E}\left[\left(\frac{n}{\chi_{n}^2}\right)^{2}\right]
      \leq \frac{C}{n^{2}},
  \end{align*}
  where the inverse chi-squared distribution has finite moment for $n$ large
  enough and $C$ is some constant related to the moments of the standard normal
  distribution and the inverse chi-squared distribution.  By the Borel-Cantelli
  lemma, we conclude the result.

\end{proof}


\begin{proof}[\textsc{Proof of Proposition \ref{P:eqw}}]  
 For the equally weighted portfolio $w = \frac{1}{\nv}1_{\nv}$ 
using
      $\eb_{\rho^*_{\nv}}$,
      \begin{align*}
        \scrR_{w} 
          & = \frac{ 
              \hat\sigma^2_{\rho^*_{\nv}} N\gamma_{\eb_{\rho^*_{\nv}}, z}^2
                 + \hat\delta^2\frac{1}{\nv} 
            }{
              \sigma^2 N\gamma_{\beta, z}^2 + \delta^2\frac{1}{\nv}
            } \\
          & \sas \frac{ \hat\sigma^2_{\rho^*_{\nv}} \gamma_{\eb_{\rho^*_{\nv}}, z}^2 }
              { \sigma^2 \gamma_{\beta, z}^2 },
      \end{align*}
      where $\hat\sigma^2_{\rho^*_{\nv}}  =
      \left(\frac{\gamma_{\eb, z}}{\gamma_{\eb_{\rho^*_{\nv}}, z}}\right)^2
      \hat\sigma^2$ is the corrected factor variance.  
Using Theorem \ref{T:bias}, we get,
      \begin{align*}
        \scrR_{w}
          \sas \frac{ \sigma^2 \rmtchisq \rmtdeflate^2 
                \rmtdeflate^{-2}\gamma_{\beta, z}^2 }
              { \sigma^2 \gamma_{\beta, z}^2 }
          \sas \rmtchisq.
      \end{align*}
\end{proof}

%% file: appendix/asymptotics.tex

Throughout, we assume Assumption \ref{a:vol}. 
The estimates $\p{\fvole,\fee,
\svole}$, are general (but with $\|\fee\|=1$) and not necessarily 
from PCA. In our setting,  formula $\req{eq:clarke}$
for the minimum variance portfolio 
simplifies to (cf. \ci{clarke2011})
\begin{align*}\label{eq:mvls}
   \optw \propto \bthr z  - \fe
   \hspace{8pt} \text{where} \hspace{8pt}
   \bthr = \frac{ \vf + \sv }
   {\vf  \gamma_{\beta, z}} \pt .
\end{align*}
Analogously,  $\minw$ and $\ebthr$ are defined in
terms of the estimates $\p{\fvole, \fee, \svole}$.  
The normalizing factor for the portolio weights that ensures
$1_{\nv}^\top \optw = 1$ is given by
\begin{align}
  W = \sqrt{\nv}\p{ \bthr - \gam{\fe}{z}} \pt .
\end{align}
We develop $(\nv \upto \infty)$-asymtotics for the tracking error
$\te$ and variance ratio $\scrR$ in equations 
$\req{te}$--$\req{var-ratio}$ for our simple one-factor model. 
It is easy to see that
\begin{align}
  \scrR_{\minw} &= 
  \frac{\fve ( \fee^\top \minw)^2 +  \sve \|\minw\|_2^2}
       { \fv ( \fe^\top \minw)^2 + \sv  \|\minw\|_2^2 } \pt , \\
  \te^2_{\minw} &=  \vf \pt ( \fe^\top \minw - \fe^\top \optw)^2
  + \sv \| \optw - \minw\|_2^2 \pt . \label{eq:teterms}
\end{align}
Note that all quantities involved depend on $\nv$. We suppress
this dependence to ease the notation. Recall $\err$ in 
$\req{error}$ defined as
\begin{align*}
  \err &= \frac{\gam{\fe}{z}-\gam{\fe}{\fee}\gam{\fee}{z}}
   {1 - \gam{\fee}{z}^2} \pt .
\end{align*}

\begin{proposition}[Asymptotics] 
Suppose Assumption \ref{a:vol} holds. Let
$\sfv = \fv/\nv$ and $\sfve = \fve/\nv$  and assume 
that $\p{\sfvole, \svole}$ are bounded in $\nv$.
\label{P:asymptotics} 
\begin{enumerate}[(i)]
\item Suppose that $\sup_\nv \gam{\fe}{z} < 1$ and 
$\sup_\nv \gam{\fee}{z} < 1$. Then,
\begin{equation}
\hspace{0.32in} 
\begin{aligned}
\scrR_{\minw} &\sim  \frac{\sve}
  {\sv + {\sfv \pt \err^2}\nv \sin^2 {\theta_{\fee,z}}} \\
\te^2_{\minw} &\sim \sfv \err^2 + 
  \sv \nv^{-1} \frac{ \p{\gam{\fee}{z}^2-\gam{\fe}{z}^2}}
  {\p{1 - \gam{\fe}{z}^2}\p{1 - \gam{\fee}{z}^2}}
\end{aligned} 
\hspace{0.32in} \p{\nv \upto \infty}.
\end{equation}
\item Suppose that $\sup_\nv \gam{\fe}{z} < 1$ and 
$\gam{\fee}{z} = 1$ eventually in $\nv$. Then,
\begin{equation}
\hspace{0.32in} 
\begin{aligned}
  \scrR_{\minw} &\sim
  \frac{ \sfve }
       { \sfv \gam{\fe}{z}^2  }  \\
  \te^2_{\minw} &\sim  \sfv \gam{\fe}{z}^2 
\end{aligned} 
\hspace{0.32in} \p{\nv \upto \infty}.
\end{equation}
\item Suppose that $\gam{\fe}{z} = 1$ eventually in $\nv$
and $\sup_\nv \gam{\fee}{z} < 1$. Then,
\begin{equation}
\hspace{0.32in} 
\begin{aligned}
  \scrR_{\minw} &\sim  \nv^{-1} \pa{ \frac{\sve /\sfv}
  {\sin^2 \theta_{\fee,z} } } \\
  \te^2_{\minw} &=  \nv^{-1} \pa{
  \frac{\sv \gam{\fee}{z}^2 }{\sin^2 \theta_{\fee,z}}  }
\end{aligned}
\hspace{0.32in} \p{\nv \upto \infty}.
\end{equation}
\item Suppose that both $\gam{\fe}{z} = 1$ and 
$\gam{\fee}{z} = 1$ eventually in $\nv$. Then,
$\te^2_{\minw} = 0$ eventually in $\nv$
and $\scrR_{\minw} \sim \fve/\fv$  as $\nv \upto \infty$.
\end{enumerate}
\end{proposition}

\begin{prove}
  All claims follow from the collection of Lemmas below.
\end{prove}

\vspace{0.16in}

It is not difficult to show the following identities
in our setting.
\begin{equation}
\begin{aligned}
  \fv \pt (\fe^\top \optw)^2
  &= \sfv
  \left( \frac{\sv \gam{\fe}{z} }
  {\sv + \fv \p{1- \gam{\fe}{z}^2} }\right)^2  \\
\fv \pt (\fe^\top \minw)^2
  &= \sfv \left( 
    \frac{\gam{\fe}{z} \sve + \fve 
     \p{\gam{\fe}{z}-\gam{\fe}{\fee}\gam{\fee}{z}}}
   {\sve + \fve  \p{ 1 - \gam{\fee}{z}^2}} \right)^2  \\
\optw^\top \optw 
    &= \frac{(\fv+\sv)^2 - \p{\fv + 2 \sv} 
   \fv \gam{\fe}{z}^2}
   {\nv (\sv + \fv\p{1 -  \gam{\fe}{z}^2})^2 }  \\
\minw^\top \optw &= \frac{\sv + \fv
   - \fv \gam{\fe}{z} E }
   { \p{\sv + \fv \p{1-\gam{\fe}{z}^2}} \nv} 
\\ E &=  \frac{\p{\sve +  \fve} \gam{\fe}{z} 
- \vfe \gam{\fee}{z}\gam{\fee}{\fe} }
{\sve + \fve \p{1-\gam{\fee}{z}^2}  }
\end{aligned}
\end{equation}
These are suffiecient to prove the following Lemmas. As
in Propositon \ref{P:asymptotics} we set 
$\sfv = \fv/\nv$ and $\sfve = \fve/\nv$  and assume 
that $\p{\sfvole, \svole}$ are bounded in $\nv$.

\begin{lemma}[True portfolio variance due to factor]
\label{L:tpv}
When $\sup_{\nv} \gam{\fe}{z} < 1$ and 
$\sup_\nv \gam{\fee}{z} < 1$, the factor components of 
the true variance of $\optw$ and $\minw$  satisfy
\begin{equation} \label{eq:fvarasymp}
\begin{aligned}
  \fv \pt (\fe^\top \optw)^2
  &\sim \frac{\svol^4}{\sfv  \nv^2} 
  \left( \frac{\gam{\fe}{z} }{1 - \gam{\fe}{z}^2} \right)^2 \\ 
  \fv \pt (\fe^\top \minw)^2
  &\sim \sfv  \err^2  
\end{aligned}
\hspace{0.32in} \p{\nv \upto \infty}.
\end{equation}
\end{lemma}

\begin{lemma}[Porfolio weights]\label{L:pw}
Suppose $\sup_{\nv} \gam{\fe}{z} < 1$ and 
$\sup_\nv \gam{\fee}{z} < 1$. 
\begin{equation}
\begin{aligned}
  \optw^\top \optw 
   &\sim \nv^{-1} 
   \pa{ \frac{1}{1 - \gam{\fe}{z}^2} } \\
   \minw^\top \optw 
   &\sim  \nv^{-1} \pa{
   \frac{ 1 - \gam{\fe}{z} \scrE }
   { 1 - \gam{\fe}{z}^2 }}
\end{aligned}
\hspace{0.32in} \p{\nv \upto \infty}.
\end{equation}
\end{lemma}
